\documentclass[%
 reprint,
 amsmath,amssymb,
 aps,
]{revtex4-2}

\usepackage{ulem}
\usepackage{graphicx}
\usepackage{dcolumn}
\usepackage{bm}
\usepackage{physics}
\usepackage{amsmath, amsfonts, dsfont}
\usepackage{siunitx}
\usepackage{accents}
\usepackage{verbatim}
\usepackage{xcolor} 
\usepackage{hyperref}

\DeclareSIUnit\gauss{G}
\DeclareSIUnit\atom{at.}

\newcommand{\sket}[1]{\left.\left| #1 \right\rangle\!\right\rangle}

\newlength{\dhatheight}
\newcommand{\dhat}[1]{%
    \settoheight{\dhatheight}{\ensuremath{\hat{#1}}}%
    \addtolength{\dhatheight}{-0.25ex}%
    \hat{\vphantom{\rule{1pt}{\dhatheight}}%
    \smash{\hat{#1}}}}

\begin{document}

\preprint{APS/123-QED}

\title{Quantum Process Tomography of a Thermal Alkali-Metal Vapor}

\newcommand{\JUAddress}{Marian Smoluchowski Institute of Physics, Jagiellonian University in Krak\'ow, 30-348 Krak\'ow, Poland}
\newcommand{\JUDSAddress}{Doctoral School of Exact and Natural Sciences, Jagiellonian University in Krak\'ow, 30-348 Krak\'ow, Poland}
\newcommand{\UAMAddress}{Institute of Spintronics and Quantum Information, Faculty of Physics, Adam Mickiewicz University, 61-614 Pozna\'n, Poland}
\newcommand{\HUaddress}{Department of Physics, Harvard University, Cambridge, MA 02138, USA}
\newcommand{\ZJUaddress}{School of Physics, Zhejiang University, Hangzhou 310027, China}

\author{Yujie Sun}
\affiliation{\JUAddress}
\affiliation{\ZJUaddress}

\author{Marek Kopciuch}
\email{marek.kopciuch@amu.edu.pl}
\affiliation{\UAMAddress}
\affiliation{\JUAddress}

\author{Arash Dezhang Fard}
\affiliation{\JUAddress}
\affiliation{\JUDSAddress}

\author{Szymon Pustelny}
\email{szymon.pustelny@uj.edu.pl}
\affiliation{\JUAddress}
\affiliation{\HUaddress}

\date{\today}

\begin{abstract}
    Characterizing the open-system dynamics of multilevel quantum systems (qudits) remains a fundamental challenge due to ensemble inhomogeneities and complex environmental interactions. Here, we introduce a computationally efficient quantum process tomography framework that reconstructs the Liouvillian dynamics of a thermal $^{87}$Rb qutrit ensemble directly in the Bloch-Fano representation. By combining maximum likelihood estimation with post-hoc spectral regularization, our protocol extracts physically admissible, completely positive and trace-preserving maps without repeated numerical integration of the master equation. We rigorously justify selecting the principal branch for the matrix logarithm by demonstrating that experimental eigenvalue phases remain strictly bounded within $[-0.35,0.35]$ radians, avoiding branch-cut ambiguities. The method is validated across relaxation-driven, static-field, and time-dependent regimes, resolving overlapping control signals and subtle dissipative mechanisms such as AC Stark shifts. Our approach establishes a scalable route for generator-level characterization of ambient qudit systems, enabling noise-aware control and precise benchmarking for atomic sensors and simulators.
\end{abstract}

\maketitle

\section{Introduction}

Ranging from quantum sensing \cite{degen2017quantum,giovannetti2011advances} to quantum computation \cite{nielsen2010quantum}, the precise characterization of quantum processes is of crucial importance to realize the full potential of quantum technologies \cite{lobino2008complete, mohseni2008quantum}. In this context, it is important to note that many of the most versatile and accessible quantum platforms, including neutral atoms \cite{keesling2019quantum,ebadi2021quantum,chaudhury2007quantum, saffman2010quantum}, trapped ions \cite{ben2020direct,figgatt2019parallel,nam2020ground}, color centers in diamond \cite{howard2006quantum}, linear optics \cite{zhou2015quantum,wang2023generalized,wang2024polarization}, and superconducting circuits \cite{arute2019quantum,samach2022lindblad}, are inherently open systems whose dynamics involve an interplay between coherent evolution and complex environmental interactions. Moreover, extending quantum control beyond two-level systems (qubits) to multi-level systems (qudits) \cite{thew2002qudit,wang2020qudits} (e.g., qutrits \cite{klimov2003qutrit}) expands the accessible Hilbert space and enhances the capabilities of quantum protocols. However, fully characterizing the evolution of these higher-dimensional systems remains challenging in realistic ambient environments, where decoherence and control imperfections substantially affect the dynamics.

Quantum Process Tomography (QPT) \cite{d2001quantum, white2022non, ahmed2023gradient, di2024fourier} provides a route to reconstruct the Liouvillian superoperator \cite{buvzek1998reconstruction} and to separate coherent and dissipative contributions to the evolution \cite{hayden2022canonical}. 
Although QPT has been extensively developed for qubits, its realization in higher-dimensional systems (qudits) remains experimentally demanding.
In particular, in thermal atomic vapors \cite{finkelstein2023practical,li2022spin}, such difficulties stem from complex interactions, principally linear Zeeman splitting of the ground-state energy sublevels (degeneracy of level splittings), environmental noise, and intrinsic inhomogeneous characteristics of the medium, including Doppler-induced broadening and spatially dependent atomic dynamics \cite{auzinsh2010optically}. 

The dynamics of open quantum systems are commonly described within the Lindblad framework \cite{lindblad1976generators,gorini1976completely} under the Born-Markov approximation \cite{bloch1957generalized,redfield1957theory,davies1974markovian,singh2024pointer}. In conventional QPT, one reconstructs the generator by measuring the system at discrete times, numerically propagating a trial master equation, and iteratively updating the Liouvillian to match the observed trajectories \cite{samach2022lindblad}. While broadly applicable, this strategy faces two major limitations in practice: repeated numerical propagation becomes increasingly costly in larger Hilbert spaces, and the formalism is typically not well suited to reconstructing genuinely time-dependent generators. Recent approaches based on continuous weak measurements \cite{siva2023time}, machine learning \cite{han2021tomography}, and Liouvillian-level tomography \cite{aguiar2025quantum} have begun to address aspects of time dependence and more general, strongly non-Markovian dynamics. However, these advancements generally rely on computationally intensive optimization or repeated dynamical inference. This leaves scalability a central challenge for higher-dimensional qudit systems.

Here, we introduce a QPT framework that addresses both limitations. By working in the generalized Pauli representation, we obtain a linear description of an effective three-level open quantum system that avoids repeated numerical propagation of the master equation during optimization and naturally extends to time-dependent Liouvillians. We experimentally validate the method on the $f=1$ ground-state manifold of a thermal $^{87}$Rb vapor ensemble, achieving high-fidelity reconstruction of qutrit Liouvillians constrained to physically admissible, completely positive and trace-preserving maps. Crucially, the representation-independent nature of this protocol means its utility extends significantly beyond atomic vapors; the framework can be seamlessly adapted to minimize gate characterization overhead in other leading multi-level quantum platforms, such as trapped-ion qudits and superconducting transmon arrays. Ultimately, this establishes a route toward generator-level characterization of ambient multilevel quantum systems.

\section{Principles of quantum process tomography\label{sec:Principles}}

Our goal is to reconstruct the evolution superoperator $\dhat{P}(t)$ that governs the observed open-system dynamics of the qutrit ensemble. Under the Born--Markov approximation, the density operator $\hat{\rho}(t)$ evolves according to the Lindblad master equation,
\begin{equation}
\frac{d}{dt}\hat{\rho}(t)=\mathcal{L}(t)\!\left [\hat{\rho}(t)\right ],
\label{eq:lindblad_compact}
\end{equation}
where $\mathcal{L}(t)$ is the Liouvillian superoperator containing both coherent and dissipative contributions to the dynamics. In the Bloch--Fano representation, the density operator is mapped onto a real $d^2$-dimensional vector $\sket{\rho(t)}$, and the superoperator $\mathcal{L}(t)$ is represented by a $d^2\times d^2$ matrix, denoted by $\dhat{L}(t)$. Thus, the evolution becomes a linear transformation in operator space,
\begin{equation}
\frac{d}{dt}\sket{\rho(t)} = \dhat{L}(t)\sket{\rho(t)},
\end{equation}
with a formal solution
\begin{equation}
\sket{\rho(t)} = \dhat{P}(t)\sket{\rho(0)} .
\label{eq:processmap}
\end{equation}

This linear representation is central to our approach. Rather than relying on computationally expensive algorithms to repeatedly propagate the master equation during fitting, we reconstruct the process directly from experimentally measured input and output state vectors. The detailed construction of the Bloch--Fano basis and the corresponding vectorization procedure are given in Supplementary Information (SI).

To determine $\dhat{P}(t)$ experimentally, we prepare an informationally complete set of initial states spanning the target manifold and reconstruct both the inputs and their corresponding outputs after evolution. In the Bloch--Fano representation, the input and output state sets form the input matrix $\mathbb{M}^{(i)}$ and the output matrix $\mathbb{M}^{(o)}$
\begin{equation}
\mathbb{M}^{(i,o)}=\left(\sket{\rho^{(i,o)}_1},\dots,\sket{\rho^{(i,o)}_N}\right),
\end{equation}
which are related through
\begin{equation}
\mathbb{M}^{(o)}=\dhat{P}(t)\mathbb{M}^{(i)}.
\label{eq:inputoutput}
\end{equation}
For an informationally complete set, this relation allows direct reconstruction of the process matrix. In our implementation, we use an overcomplete set of fifteen specific initial states to improve numerical stability and robustness against experimental noise; accordingly, we reconstruct $\dhat{P}(t)$ from the symmetrized form of Eq.~(\ref{eq:inputoutput}),
\begin{equation}
    \dhat{P}(t)=\mathbb{M}^{(o)}\left(\mathbb{M}^{(i)}\right)^{-1}.
    \label{eq:linearinv}
\end{equation}
Thus, QPT reduces to determining a linear map in the operator space spanned by the experimentally prepared states. 

Equation~\eqref{eq:linearinv} yields the raw, unconstrained process matrix $\dhat{P}(t)$. Because empirical data inevitably contains statistical noise and state-preparation-and-measurement (SPAM) errors, this raw algebraic inversion is prone to static gauge ambiguities and is not guaranteed to represent a physically valid, completely positive and trace-preserving (CPTP) map. Therefore, this linear inversion serves only as the first step of our protocol. The corrected matrices are subsequently subjected to a Maximum Likelihood Estimation (MLE) to fit the continuous Liouvillian generator. Crucially, the extraction of the generator relies on the principal branch of the matrix logarithm, which is physically justified here as the empirical eigenvalue phases remain strictly bounded, avoiding branch cut discontinuities. Finally, we apply a post-hoc spectral regularization to the reconstructed generator to strictly enforce physical admissibility and CPTP compliance.

\begin{figure*}[htbp]
    \centering
    \includegraphics[width=2\columnwidth]{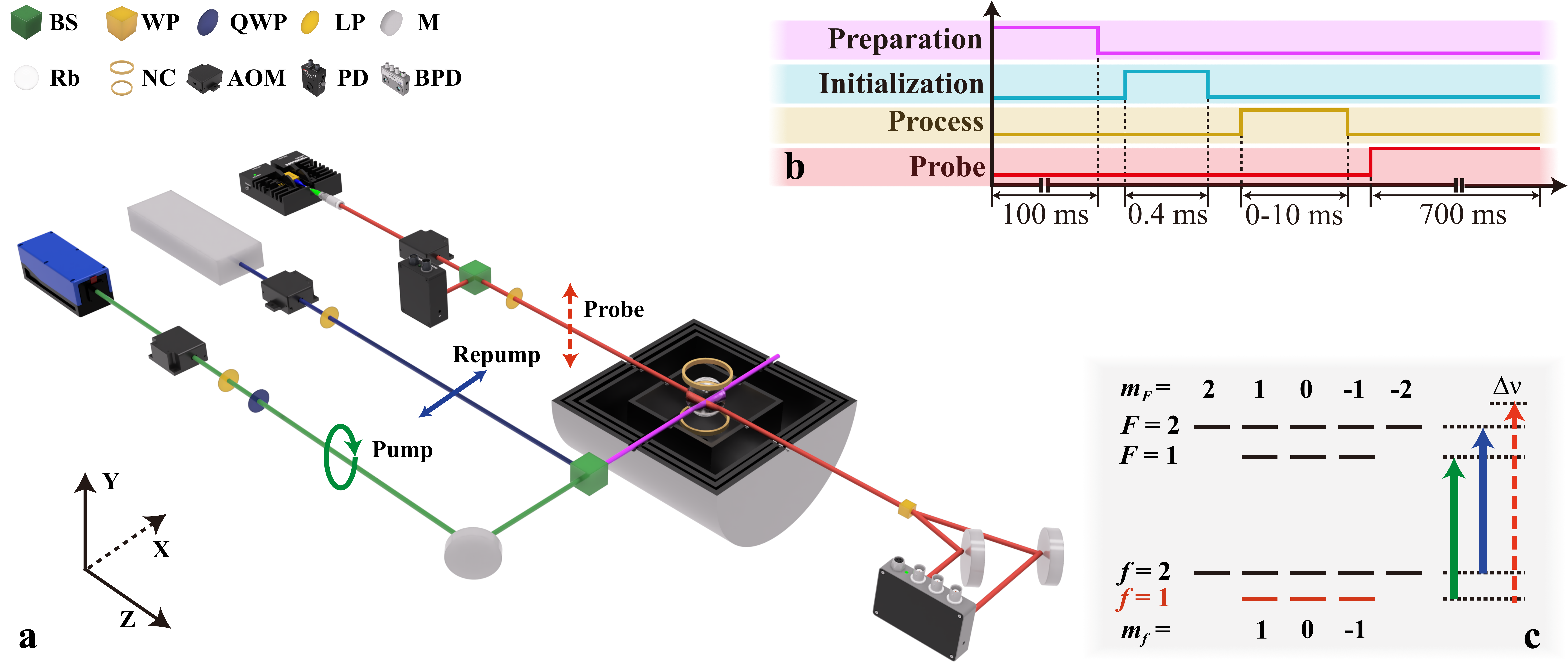}
    \caption{Experimental scheme for quantum process tomography of a thermal $^{87}$Rb vapor qutrit. Shown are the optical and magnetic-field setup (\textbf{a}), the experimental sequence (\textbf{b}), and the relevant energy-level structure of the $^{87}$Rb D$_1$ line (\textbf{c}). The protocol comprises four stages: preparation of a spin-polarized state, initialization of one of 15 informationally complete input states, application of the target process, and probe-based readout of the output state through polarization rotation. Reconstructing the full set of input and output states by quantum state tomography \cite{Kopciuch2024} yields the corresponding process matrix. BS, beam splitter; WP, Wollaston prism; QWP, quarter-wave plate; LP, linear polarizer; M, mirror; Rb, paraffin-coated $^{87}$Rb vapor cell; NC, nonlinear-control coils; AOM, acousto-optic modulator; PD, photodiode; BPD, balanced photodetector.
    }
    \label{fig:Setup and time sequence}
\end{figure*}

A conceptual overview of our experimental apparatus and the tomographic sequence is presented in Fig.~\ref{fig:Setup and time sequence}. Full details regarding the optical configuration, state initialization, continuous readout, and the rigorous SPAM calibration procedures are provided in the Methods section and SI.

\section{Experimental reconstruction of qutrit Liouvillians\label{sec:Liouvillian}}

\begin{figure}[htbp]
    \centering
    \includegraphics[width= 0.95\linewidth]{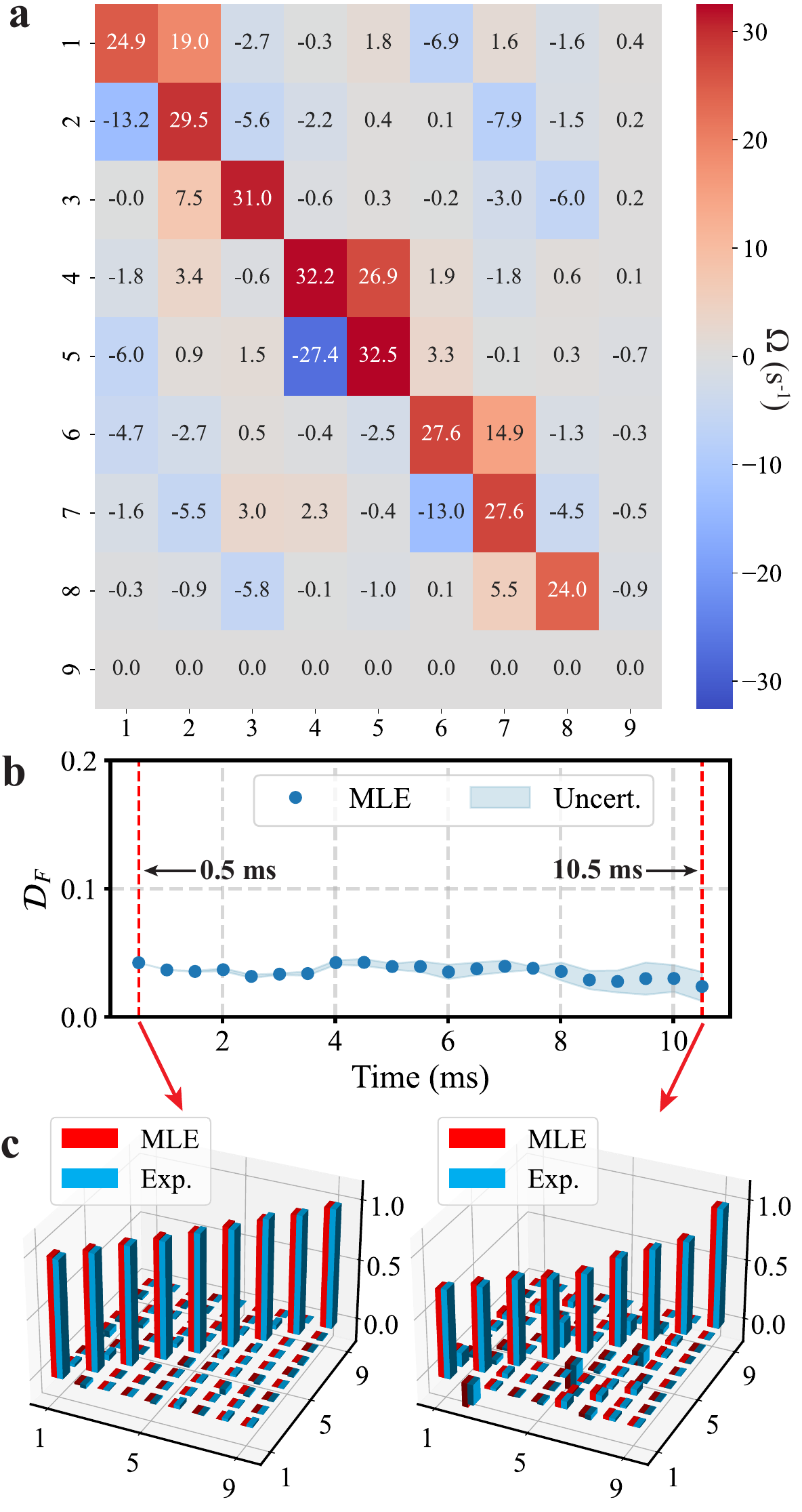}
    \caption{(a) Reconstruction of the effective total relaxation superoperator $\dhat{R}_{T}$ via the MLE protocol. (b) The relative error $\mathcal{D}_{F}$ with the uncertainty (Uncert.) quantifies the difference between the process matrix $\dhat{P}(t)$ obtained from experimental measurements and the simulation results via MLE over time. (c) The bar charts visually compare the experimental process matrix $\dhat{P}_{\rm Exp.}$ (blue bar) and simulation via MLE $\dhat{P}_{\rm MLE}$ (red bar) at $t = 0.5, \SI{10.5}{\milli\second} $, and illustrate their deviations respectively.
    }
    \label{fig: superoperator relaxation}
\end{figure}

With the calibrated process matrices in hand, we now reconstruct the generators governing the dynamics of the thermal $^{87}$Rb vapor qutrit. Throughout this section, the evolution is interpreted as the sum of a controlled coherent contribution and an effective time-independent relaxation term,
\begin{equation}
\dhat{L}(t) = -i\left[\dhat{H}_{C}(t)+\dhat{H}_{R}\right]-\dhat{R}
= -i\dhat{H}_{C}(t)-\dhat{R}_{T},
\label{eq:Lsplit}
\end{equation}
where $\dhat{H}_{C}(t)$ denotes the applied control Hamiltonian superoperator, $\dhat{H}_{R}$ is a residual uncontrolled Hamiltonian superoperator, and $\dhat{R}$ describes intrinsic dissipation. For reconstruction, we absorb the static residual Hamiltonian superoperator into a time-independent effective total relaxation superoperator $\dhat{R}_{T}$.

In a simplified scenario where the Liouvillian is assumed time-independent, a measurement at a single evolution time $t'$ can provide a direct algebraic estimate of the generator: $\dhat{L}=\frac{1}{t'}\log\!\left[\dhat{P}(t')\right]$. Because the complex matrix logarithm is a multi-branched function, branch selection must be rigorously justified. Empirical analysis of our reconstructed process matrices demonstrates that the eigenvalue phases are strictly confined to the interval $[-0.35, 0.35]$ radians. Consequently, the phase evolution does not approach the $\pm \pi$ discontinuity, ensuring that the $2\pi$ rotational ambiguity is physically inaccessible. This validates the use of the principal branch.

To achieve the highest-fidelity reconstruction for such time-independent dynamics by enforcing physical constraints, and to establish a rigorous benchmark for evaluating our direct algebraic inversion, we concurrently employ MLE to minimize the discrepancy over the entire dataset:
\begin{equation}
\Delta^2(\dhat{L})
=
\sum_n
\left\|
\exp(\dhat{L}t_n)-\dhat{P}(t_n)
\right\|_F^2,
\end{equation}
where $\lVert\cdot\rVert_F$ denotes the Frobenius norm. 

To assess the quality of Liouvillian reconstruction and to compare process matrices, we employ the normalized Frobenius distance~\cite{boulant2003robust,zhang2012recursive} defined as
\begin{equation}
    \mathcal{D}_F (\dhat{A}, \dhat{B}) = \frac{ \norm{\dhat{A} - \dhat{B}}_F }{ \norm{\dhat{B}}_F },
    \label{eq:relative_error}
\end{equation} 
where $\dhat{A}$ and $\dhat{B}$ are superoperators corresponding to either measured and estimated process matrices or reconstructed and theoretical Liouvillian superoperators.

\subsection{Estimation of the dissipator \texorpdfstring{$\dhat{R}_{T}$}{R}}\label{sec:estimation_of_dissipator}

We first determine the effective total relaxation superoperator $\dhat{R}_{T}$ in a dedicated reference experiment with all controllable external fields switched off. Under these conditions, the measured dynamics are governed solely by the intrinsic non-unitary evolution of the atomic ensemble, so that $\dhat{R}_{T}$ provides the dissipative baseline for all subsequent Hamiltonian reconstruction. Once this background is established, it can be held fixed in later measurements to isolate the controlled coherent contribution $\dhat{H}_{C}(t)$.

Figure~\ref{fig: superoperator relaxation}(a) shows the reconstructed relaxation superoperator obtained from process matrices measured between 0.5 and \SI{10.5}{\milli\second}, using the MLE described with $\dhat{H}_{C}=0$. The resulting generator is consistent with trace preservation: because the total atomic population is conserved, the ninth row of $\dhat{R}_{T}$ must vanish, which is strictly confirmed by the reconstructed matrix.

At the level of a first approximation, the reconstructed superoperator is dominated by an isotropic relaxation component, as commonly assumed for thermal atoms in paraffin-coated vapor cells after compensation of residual fields and gradients. In this limit, relaxation would appear approximately diagonal in the Bloch--Fano representation, with an isotropic rate of about $28.7(24)\,\si{\per\second}$. However, Fig.~\ref{fig: superoperator relaxation}(a) also reveals several pronounced off-diagonal elements, indicating that a purely isotropic model is insufficient to capture the measured dissipative dynamics.

Motivated by this structure, we expand our theoretical model to include three distinct mechanisms governing $\dhat{R}_T$. The first involves coherent contributions induced by residual magnetic fields together with the AC Stark shift arising from the leakage of the strong repump beam. These induce weak coherent precession with residual Larmor frequencies $\Omega_{L}/2\pi=\{ -0.6828(25), 0.1142(86), 2.240(34)\}$~Hz and an AC Stark-shift frequency $\Omega_\text{AC}/2\pi=0.056(16)$~Hz. The second mechanism is the isotropic relaxation occurring at a rate $\gamma_i=14.84(15)$ $\rm{s}^{-1}$. The third captures decoherence induced by magnetic-field inhomogeneities, modeled via three independent Lindblad operators along the spatial axes. The corresponding dephasing rates are $\gamma_k=\{6.7(12),6.1(12),5.60(11)\}$~$\rm{s}^{-1}$. As detailed in the SI, this multi-component decomposition accurately captures the vapor cell's internal dynamics. Crucially, the ability to resolve these subtle off-diagonal contributions underscores the diagnostic power of the Bloch--Fano framework over simplistic uniform decay models. Unless otherwise specified, all reported values represent total standard uncertainties ($u_{total}$), incorporating statistical variance and algorithmic bias derived from the spectral regularization process.

Using the reconstructed $\dhat{R}_{T}$, we simulate the corresponding process evolution according to $\dhat{P}_{\mathrm{MLE}}(t)=\exp(-\dhat{R}_{T}t)$ and compare it directly with the experimentally measured process matrices. As shown in Fig.~\ref{fig: superoperator relaxation}(b), the relative discrepancy remains small throughout the evolution window, with a maximum normalized Frobenius distance of $0.04929(16)$. Representative comparisons at short and long evolution times [Fig.~\ref{fig: superoperator relaxation}(c)] further show close agreement between the measured process matrices and those predicted from the reconstructed dissipator, confirming that $\dhat{R}_{T}$ provides an accurate and physically consistent description of the intrinsic relaxation dynamics. 

\subsection{Time-independent Hamiltonian\label{sec:Time-independent Hamiltonian}}
\begin{figure}
    \centering
    \includegraphics[width=1\linewidth]{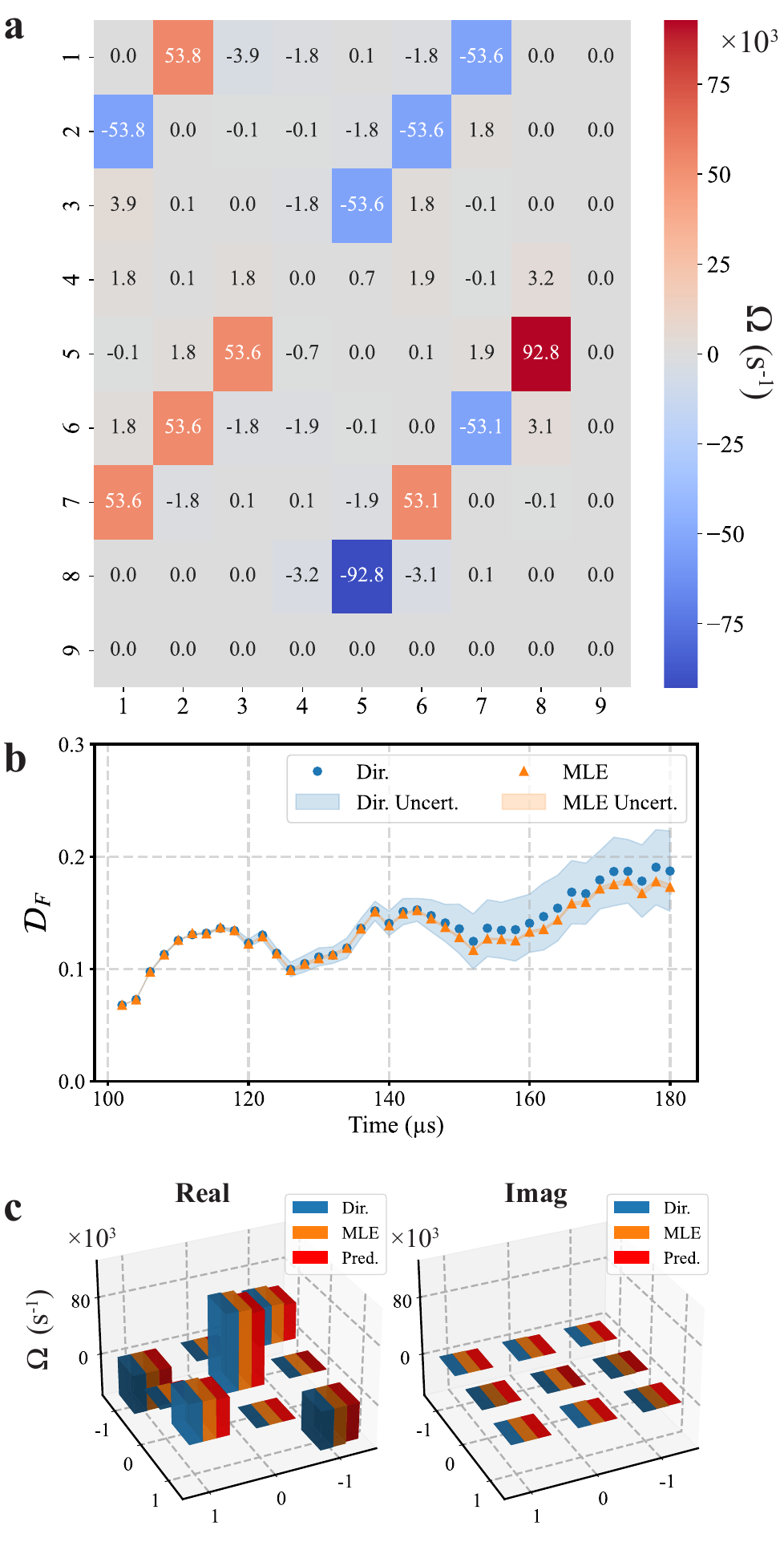}
    \caption{(a) Hamiltonian-superoperator reconstruction using the MLE approach. (b) Normalized Frobenius distance between the experimentally measured process superoperators and those generated from the estimated $\dhat{H}_C$ using direct reconstruction (blue dots) and MLE (orange triangles). The blue (direct uncertainties) and orange (MLE uncertainties) shaded regions represent the associated uncertainties, respectively. (c) Comparison of the real and imaginary parts of Hamiltonian operators. The red bars show the predicted Hamiltonian (Pred.) based on the known setup parameters of the coil geometry and system response~\cite{dezhang2025isolating}, while the orange (MLE) and blue (direct reconstruction) ones represent the Hamiltonians extracted from the estimated superoperators. The relative error, quantified by the normalized Frobenius distance between the predicted and MLE-based Hamiltonian, is $\mathcal{D}_F^\text{MLE} = 0.05657(87)$, while for direct reconstruction it is $\mathcal{D}_F^\text{Dir} = 0.068(15)$.}
    \label{fig:NonliearY_total}
\end{figure}

Having established the intrinsic relaxation background, we next reconstruct the controlled Hamiltonian in the time-independent regime. This case provides a stringent validation of the method, because the applied perturbation is independently known and the reconstructed generator can therefore be compared directly with a physically calibrated reference.

As a test platform, we consider coherent evolution generated predominantly by the quadratic Zeeman effect. Experimentally, this regime is realized by applying a rapidly oscillating magnetic field along the $y$-axis with a frequency of \SI{326}{\kilo\hertz} and an amplitude exceeding \SI{25}{G}. Although the field instantaneously induces both linear and quadratic Zeeman shifts, the linear contribution averages to zero over many oscillation periods because of the alternating field sign, whereas the quadratic term accumulates coherently \cite{dezhang2025isolating}. Since the oscillation period ($\approx\SI{3}{\micro\second}$) is more than four orders of magnitude shorter than the characteristic timescale of the slow ground-state dynamics (of order $\SI{0.1}{\second}$), the resulting evolution can be treated as described by an effective time-independent Hamiltonian $\hat{H}_{C}$.

To reconstruct the corresponding control superoperator $\dhat{H}_{C}$, we use the same informationally complete set of initial states as in the relaxation experiment and measure their evolution for pulse durations between 100 and \SI{180}{\micro\second}. We then compare two reconstruction strategies: direct linear inversion, which provides a fast estimate at each evolution time but does not enforce physical constraints, and a constrained MLE reconstruction, which fits the full dataset while explicitly restricting the Hamiltonian to remain Hermitian by mapping the parameters onto the antisymmetric structure constants of the SU(3) algebra (see Section IX of the SI).

Figure~\ref{fig:NonliearY_total}(a) shows the Hamiltonian superoperator reconstructed by the MLE procedure. As expected for purely coherent evolution, the reconstructed superoperator is purely imaginary, with vanishing diagonal entries and a zero ninth row enforced by trace preservation. The remaining off-diagonal elements encode the couplings between Bloch--Fano components induced by the quadratic Zeeman interaction and reproduce the expected structure of the effective Hamiltonian.

To assess reconstruction quality, we simulate the process evolution using the previously determined $\dhat{R}_{T}$ together with the reconstructed Hamiltonian obtained from either direct inversion or MLE, and compare the resulting process matrices with the experiment. As shown in Fig.~\ref{fig:NonliearY_total}(b), both approaches reproduce the measured dynamics well, with the MLE estimate yielding a slightly smaller discrepancy. A direct comparison with the independently predicted Hamiltonian [Fig.~\ref{fig:NonliearY_total}(c)] shows similarly good agreement, again with the MLE result performing marginally better. The remaining small mismatch likely reflects weak additional decoherence induced by the spatial inhomogeneity of the driving field, which cannot be fully separated from the coherent contribution under the present experimental conditions.

\subsection{Time-dependent Hamiltonian \label{sec:time_dependant}}

\begin{figure}
    \centering
    \includegraphics[width=0.9\linewidth]{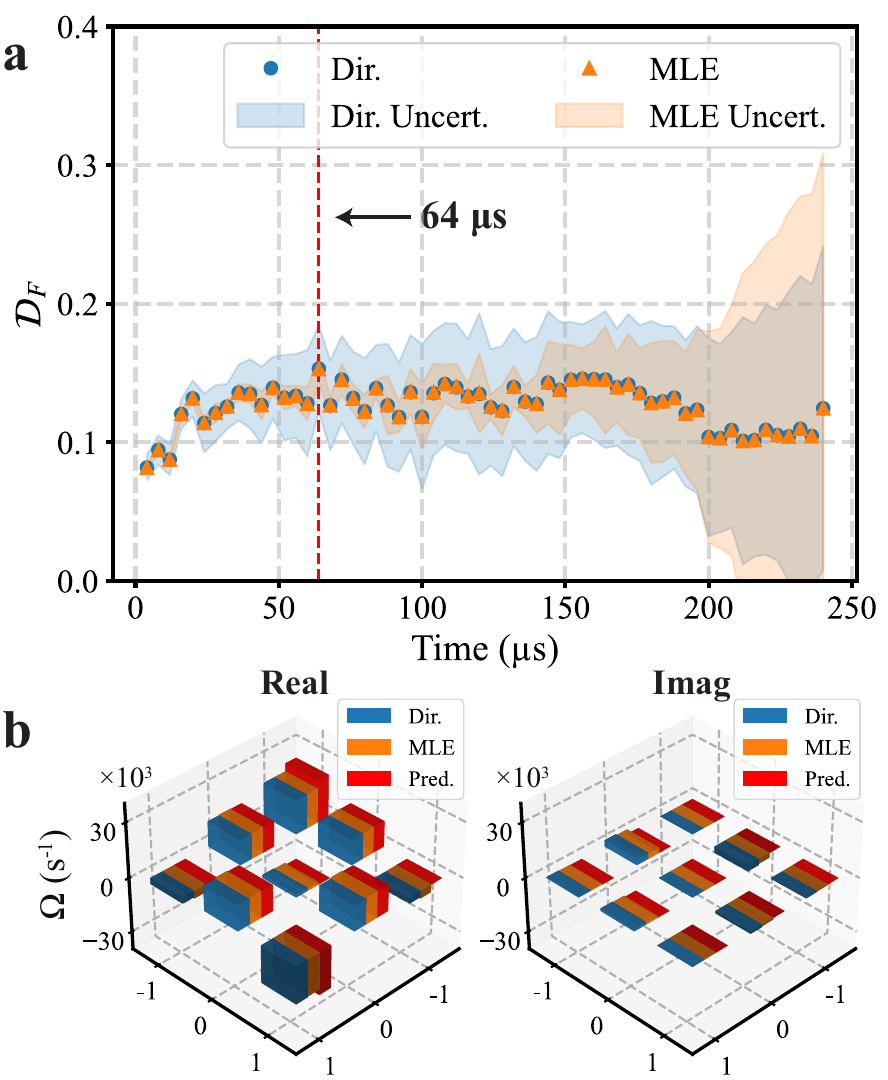}
    \caption{(a) Relative error between the experimentally measured process superoperators and those generated from the estimated time-dependent Hamiltonian $\dhat{H}_{C}(t)$ using direct reconstruction (blue dots) and MLE (orange triangles); and the corresponding uncertainties represented by the blue (direct uncertainty) and orange (MLE uncertainty) shaded regions. (b) Comparison of the real and imaginary parts of Hamiltonian operators at t =  \SI{64}{\micro\second}. The blue, orange, and red bars represent the Hamiltonian operator via direct reconstruction, MLE, and prediction, respectively. The predicted Hamiltonian is derived from known current-to-magnetic strength ratios in three directions, computed using current magnitudes at preset time points. The corresponding relative error between the predicted and MLE-based Hamiltonian superoperators is  $\mathcal{D}_F^\text{MLE} = 0.215(42)$, while for direct reconstruction it is $\mathcal{D}_F^\text{Dir} = 0.212(22)$.}
    \label{fig: Time dependent Hamiltonian}
\end{figure}

We finally consider the most demanding regime, in which the control Hamiltonian varies over time. Unlike the static case, the dynamics can no longer be described by a single time-independent generator and must instead be reconstructed over a sequence of short time intervals. Because generators at different times generally do not commute, the total evolution must be written in time-ordered form $\mathcal{T}$ rather than as a single exponential map. We therefore adopt a coarse-grained description in which the evolution is divided into intervals $\Delta t$ that remain short compared with the characteristic timescale of the applied modulation. The validity of this assumption relies on a careful selection of the time grid. The coarse-grained interval $\Delta t$ must be chosen so that it satisfies the Nyquist-Shannon sampling criterion relative to the maximum frequency component of the driving field, effectively oversampling Hamiltonian dynamics. However, we note that the ultimate bandwidth of time-dependent processes that this protocol can accurately reconstruct is fundamentally limited by the finite readout time of the probe beam. Taking into account the discretization, Eq.~\eqref{eq:processmap} can be reformulated as
\begin{equation}
    \begin{split}
        \sket{\rho(t_n)} &= \mathcal{T} \left\lbrace \prod_{k=0}^{n-1} \exp(\dhat{L}_k\,\Delta t) \right\rbrace \sket{\rho(0)} \\
        &= \mathcal{T} \left\lbrace \prod_{k=0}^{n-1} \dhat{P}_k \right\rbrace \sket{\rho(0)},
    \end{split}
\end{equation}
where $\dhat{P}_k$ is the process superoperator describing evolution from time $t_k$ to $t_{k+1}$. Measurement of state matrices at these discrete times and applying Eq.~\eqref{eq:inputoutput}, we find
\begin{equation}
    \mathbb{M}^{(n+1)} = \dhat{P}(t_n; t_{n+1})\mathbb{M}^{(n)} = \dhat{P}_{n}\mathbb{M}^{(n)}.
\end{equation}

Figure~\ref{fig: Time dependent Hamiltonian}(a) summarizes the reconstruction quality by showing the normalized Frobenius distance between the experimentally measured process superoperators and those generated from the reconstructed time-dependent Hamiltonian $\dhat{H}_C(t)$. During most of the measurement window, both direct reconstruction and MLE reproduce the observed dynamics well, indicating that the protocol remains applicable even when the control generator changes continuously over time. Crucially, by segmenting the evolution and utilizing the algebraic Bloch--Fano inversion at each discrete step, we circumvent the immense computational cost of iteratively calculating time-ordered exponentials over the entire continuous trajectory—a major scalability bottleneck in concurrent gradient-descent regression models tailored for open quantum systems.

To visualize the quality of the reconstruction during this noisy dynamical evolution, we examine the reconstructed operators at a representative time point of $t=\SI{64}{\micro\second}$. As shown in Fig.~\ref{fig: Time dependent Hamiltonian}(b), we compare the reconstructed matrices at this instant against the predicted reference Hamiltonian, which is determined from preset current magnitudes using known geometric scaling factors. The close agreement between the estimated superoperators and the theoretical target [$\mathcal{D}_F^\text{MLE} = 0.215(42)$ and $\mathcal{D}_F^\text{Dir} = 0.212(22)$] confirms that the method faithfully captures the underlying physical interaction. The residual deviations are an expected consequence of intrinsic approximations in superoperator mapping and finite sampling. Furthermore, the apparent divergence in uncertainty beyond $t\approx \SI{150}{\micro\second}$ is a natural consequence of signal depletion; as intrinsic decoherence ($\gamma_{i}= \SI{14.84(15)}{\per\second}$) damps the macroscopic atomic polarization, the diminishing signal-to-noise ratio severely exacerbates the relative error metric.

\subsection{Multiparameter estimation}

\begin{figure}[htbp]
    \centering
    \includegraphics[width=0.9\linewidth]{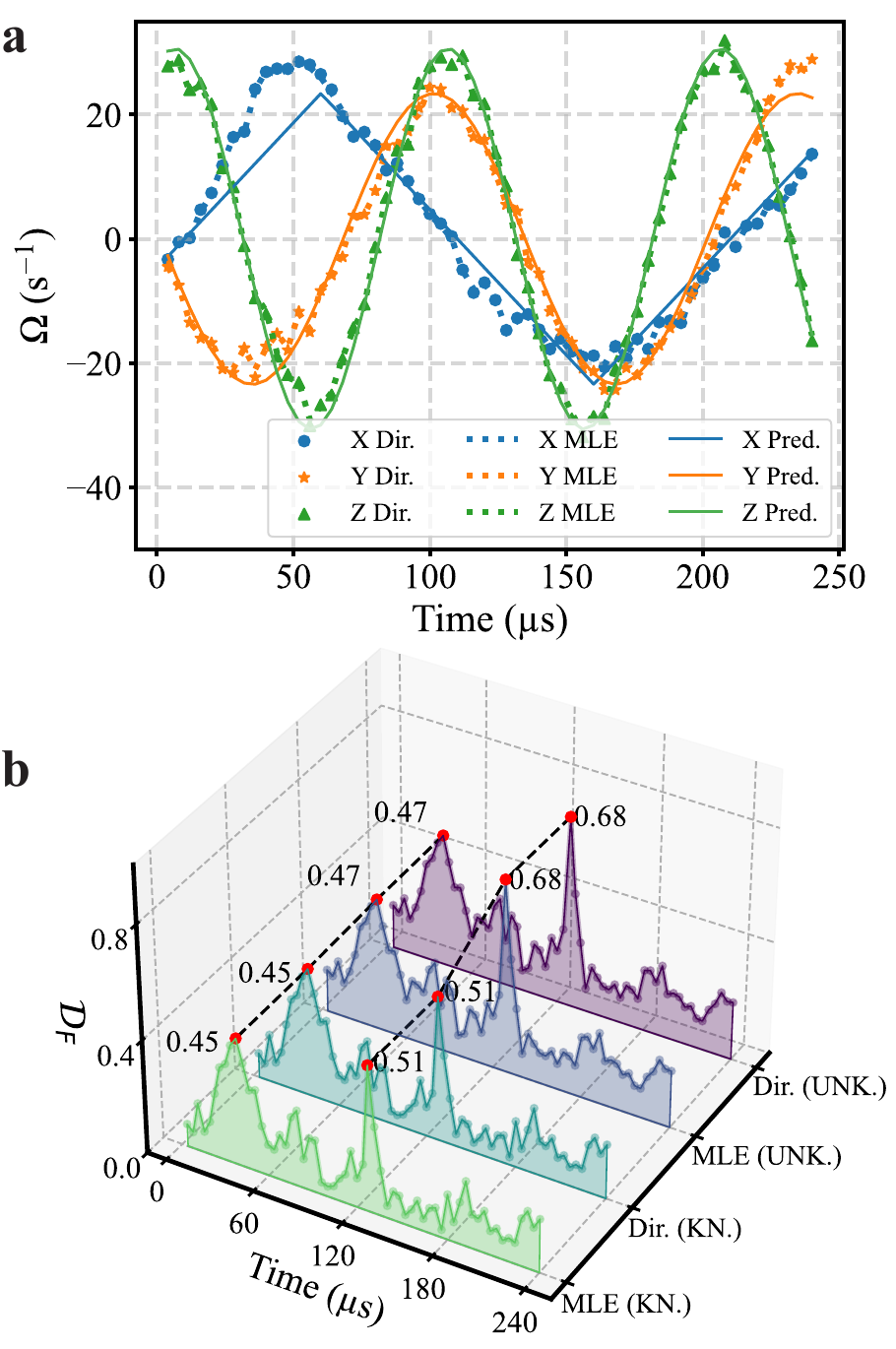}
    \caption{(a) Example of multiparameter estimation based on time-dependent Hamiltonian reconstruction. Magnetic fields are applied simultaneously along all three orthogonal directions with distinct waveforms, amplitudes, phases, and frequencies. Solid lines indicate the reference Larmor frequencies expected from the known applied magnetic fields, while dotted lines and points denote estimates obtained via MLE and direct reconstruction, respectively. (b) Relative errors between the predicted and reconstructed superoperators for direct reconstruction and MLE, with and without imposing the known Hamiltonian form constraint of Eq.~\eqref{eq:hamiltonian_model}. Purple and blue lines correspond to the unconstrained reconstruction, whereas cyan and green lines denote the constrained cases. Red markers indicate representative error values at $t=\SI{36}{\micro\second}$ and $t=\SI{124}{\micro\second}$.}
    \label{fig:time-dependent magnetic field}
\end{figure}

The time-dependent reconstruction framework can be naturally extended to the simultaneous estimation of multiple control parameters. In this setting, we assume that the functional form of the control Hamiltonian is known, while the time-dependent coefficients remain to be inferred from the data. Specifically, we consider
\begin{equation}
    \hat{H}_C(t)=\sum_k \Omega_k(t)\hat{F}_k,
    \label{eq:hamiltonian_model}
\end{equation}
where $\hat{F}_k$ denotes the angular-momentum operator along the $k$ axis and $\Omega_k(t)$ is the corresponding time-dependent Larmor frequency treated as an unknown parameter.

To estimate these parameters, we use both reconstruction strategies introduced above. In the direct approach, the time-dependent Hamiltonian is first reconstructed at each time step, and the resulting operator is then fitted by least squares to the model of Eq.~\eqref{eq:hamiltonian_model}. In the MLE approach, the same parametric form is imposed directly during optimization, so that the unknown Larmor frequencies $\Omega_k(t)$ are estimated within the constrained likelihood fit.

As a demanding test of this multiparameter protocol, we simultaneously apply three distinct time-dependent magnetic fields along the orthogonal $x$-, $y$-, and $z$-axes. The three components differ in frequency (5, 7.5, and 10~kHz), initial phase (0, $\pi$, and $\pi/2$), and waveform, with a triangular modulation along $x$ and sinusoidal modulations along $y$ and $z$. This setting probes whether the method can separate and track multiple overlapping control signals within the same driven evolution.

The results are shown in Fig.~\ref{fig:time-dependent magnetic field}(a). The parameters extracted from both MLE (dotted lines) and direct reconstruction (points) closely follow the preset reference Larmor frequencies (solid lines), demonstrating that the method can recover the distinct evolution of all three control components simultaneously. A small discrepancy is visible for the $x$ component during the initial \SI{64}{\micro\second}, which we attribute to the finite response time of the current supply. After this short transient, the reconstructed parameters converge to the expected values, confirming that the framework can faithfully capture multiple overlapping time-dependent signals.

We further examine how prior knowledge of the Hamiltonian structure affects reconstruction accuracy. As shown in Fig.~\ref{fig:time-dependent magnetic field}(b), imposing the known Hamiltonian form of Eq.~\eqref{eq:hamiltonian_model} systematically reduces the relative error of the reconstructed superoperator for both direct reconstruction and MLE. This demonstrates that physically motivated constraints derived from the underlying control model substantially improve reconstruction fidelity. The four reconstruction scenarios exhibit two concurrent error maxima. The first, at $t=\SI{36}{\micro\second}$, is mainly caused by the aforementioned transient instability of the current supply at the beginning of the pulse sequence. The second, at $t=\SI{124}{\micro\second}$, occurs when the total applied Larmor frequency approaches zero, causing the normalized error metric to amplify by the reduced magnitude of the reference signal. Together, these results show that the present framework supports robust time-resolved multiparameter estimation even in a complex driven qutrit system. The ability to track multiple time-dependent parameters in a noisy thermal environment suggests this algebraic reconstruction method can serve as a practical tool for characterizing multi-level dynamics across various quantum hardware.

\section{Conclusion\label{sec:Conclusions}}

We have introduced and experimentally demonstrated the QPT protocol that overcomes key limitations of conventional methods. By casting the Lindblad master equation in the Bloch--Fano representation, we developed a computationally efficient scheme that enables full reconstruction of time-dependent Liouvillian superoperators while greatly reducing measurement and analysis overheads. Tests on a $^{87}$Rb qutrit confirmed the accuracy and versatility of the method. Because the procedure is representation-independent, it can be extended to higher-dimensional qudits without conceptual changes.

The QPT method implemented here can be used to directly reveal Liouvillian exceptional points \cite{minganti2019quantum}, as theoretically predicted in Ref.~\cite{kopciuch2025liouvillian} for thermal alkali-metal vapors modeled as a qutrit or quartit, analogous to the application of QPT for identifying such points in few-qubit quantum circuits \cite{abo2024experimental}.

Beyond these immediate achievements, our work paves the way for several future directions. An outstanding challenge in quantum science is the characterization of non-Markovian dynamics \cite{wolf2008assessing}, where memory effects in the environment play a significant role and the standard Lindblad framework becomes inadequate~\cite{white2022non, Rivas2014Quantum, Luan2024Quantum}. Recent concurrent protocols, such as quantum Liouvillian tomography \cite{aguiar2025quantum}, address this by utilizing gradient-based regression over the derivatives of Pauli string probability distributions to capture non-Markovian effects in time-continuous quantum dynamics. While our present framework operates within the Born--Markov approximation to provide a highly efficient, regression-free algebraic inversion, its sub-millisecond time resolution means it can be adapted to evaluate non-Markovianity. By applying to successive short measurement windows, one could reconstruct the process tensor mathematical construct that fully describes environmental memory effects thereby providing direct experimental access to temporal correlations that characterize non-Markovian noise~\cite{white2022non, Rivas2014Quantum}.

A further application lies in the quantitative benchmarking of analogue quantum simulators and sensors.  Many state-of-the-art platforms, such as Rydberg-atom arrays~\cite{Browaeys2020Many}, NV centers~\cite{joas2025high}, superconducting circuits~\cite{dai2021calibration}, or trapped ions~\cite{Blatt2012Quantum}, operate in regimes where coherent and dissipative processes compete.  Reconstructing the Liouvillian under different operating conditions would allow experimentalists to validate microscopic models of their devices, identify unexpected loss channels, and refine parameter regimes for optimal performance.  Because the reconstruction is measurement-efficient, such benchmarking could be repeated throughout extended experimental runs, providing a practical tool for long-term calibration and drift compensation. Ultimately, this highlights the broad applicability and cross-disciplinary impact of our method.

Finally, the rich data produced by our method are well-suited to machine-learning analysis.  Deep-learning models have been used to accelerate state tomography~\cite{Torlai2018Neutral}, identify correlated noise~\cite{Zhang2025Learning}, and design control pulses~\cite{Bukov2018Reinforcement,Niu2021Universal}.  Training these models on time-resolved Liouvillian estimates could reveal subtle correlations, predict instabilities before they affect performance, or suggest improved operating points.  By combining efficient tomography with data-driven inference, the present approach provides a practical pathway toward the systematic characterization, calibration, and optimization of emerging quantum technologies.

\section*{Methods}

\subsection*{Experimental setup and optical control}
The experiment employs a paraffin-coated spherical vapor cell ($\SI{3.7}{\centi\meter}$ in diameter) containing an isotopically enriched sample of $^{87}\text{Rb}$ ($>99$\%). The cell is heated to $\SI{38}{\celsius}$ to optimize the signal-to-noise ratio, yielding an atomic number density of approximately $9\times10^9\text{ atoms/cm}^3$. The cell is housed within a multi-layer magnetic shield comprising three cylindrical layers of mu-metal and a single cubic layer of ferrite, providing a shielding factor of approximately $10^6$. Internal magnetic coils are used to compensate for residual fields and to apply controlled static and oscillating magnetic fields. 

Optical control relies on three independently stabilized lasers operating on the $^{87}\text{Rb}$ $D_1$ line. State initialization is performed by a pump beam generated by a Fabry-Pérot extended-cavity diode laser (ECDL) locked to the $f=1 \rightarrow F=1$ transition. The pump beam propagates along the $z$-axis and is circularly polarized using a Glan-Thompson polarizer and a quarter-wave plate. A Ti:Sapphire repump laser, locked to the $f=2 \rightarrow F=2$ transition, operates continuously to prevent population trapping in dark hyperfine states, ensuring the validity of the trace-preservation constraint. The repump beam propagates along the $z$-axis and is linearly polarized to avoid inducing spurious spin orientation. State readout is performed by an off-resonant probe beam from a distributed-feedback (DFB) diode laser, blue-detuned by 50 MHz from the $f=1 \rightarrow F=2$ transition. The probe beam propagates along the $z$-axis, is linearly polarized along the $y$-axis, and its intensity is attenuated to $\SI{10}{\micro\watt/\centi\meter\squared}$ to minimize destructive back-action on the atomic state.

\subsection*{Quantum state tomography and SPAM calibration}
Following a $\SI{100}{\milli\second}$ optical-pumping phase, the atomic ensemble is driven into one of fifteen informationally complete target states using combinations of strong oscillating and weak static magnetic fields. The evolved states are subsequently measured using Quantum State Tomography (QST) via time-resolved Faraday rotation. The polarization rotation is recorded continuously using a balanced polarimeter and isolated from technical noise using a cyclically ordered phase sequence (CYCLOPS) method. Linear inversion is then applied to reconstruct the nine-dimensional state vectors in the Bloch--Fano representation.

To mitigate SPAM distortions, we implement a zero-time calibration procedure. Because exact reconstruction at $t=0$ is experimentally inaccessible, the static SPAM contribution ($\dhat{P}_\text{SPAM}$) is estimated through linear extrapolation of the time-dependent process matrices in the matrix-logarithm representation: $\log[\dhat{P}_{exp}(t)] \approx \log(\dhat{P}_\text{SPAM}) + \dhat{L}t$. To isolate the intrinsic system dynamics, all measured process matrices are pre-multiplied by the inverse of this SPAM matrix prior to generator reconstruction.

\subsection*{Maximum likelihood estimation and spectral regularization}
Physical generators are extracted from the calibrated process matrices using an MLE that minimizes the Frobenius distance between the measured process and the simulated matrix exponential. To ensure physical admissibility, the extracted parameters are projected onto the boundary of CPTP maps. 

For Hamiltonian reconstruction, Hermiticity is guaranteed by parameterizing the superoperator using the completely antisymmetric structure constants of the SU(3) algebra. For the dissipative sector, the effective relaxation superoperator is mathematically mapped to a modified Kossakowski matrix. To eliminate non-physical decoherence rates induced by experimental noise, a post-hoc spectral regularization is applied to this matrix: negative eigenvalues are truncated to zero, and the remaining positive eigenvalues are rescaled to preserve the total trace of the dissipative sector. This yields a physically admissible, positive semi-definite GKLS dissipator. Detailed mathematical mappings between the superoperators and the SU(3) structure constants are provided in the SI.

\subsection*{Error analysis and parametric bootstrapping}
Due to the strong non-linearities of matrix inversion and logarithms, standard analytical error propagation is insufficient for calculating the combined uncertainty. Therefore, experimental uncertainties are quantified using a Monte Carlo parametric bootstrapping technique. We generate 1,000 synthetic datasets by resampling the input and output state vectors from normal distributions centered on their experimental expectation values, with standard deviations defined by their respective QST standard uncertainties. 

Projecting unphysical estimates onto the completely positive cone via asymmetric eigenvalue truncation skews the reconstructed parameter distribution, introducing an algorithmic bias between the primary MLE estimate and the ensemble mean. To account for this, the observed bias ($\Delta_{bias}$) is treated as a systematic uncertainty component. The total standard uncertainty reported for all Liouvillian parameters is calculated by combining this systematic bias in quadrature with the statistical uncertainty ($u_{stat}$) of the Monte Carlo ensemble: $u_{total} = \sqrt{\Delta_{bias}^2 + u_{stat}^2}$.

\begin{acknowledgments}
The authors thank Adam Miranowicz, Piotr Put, Rafał Demkowicz-Dobrzański, and Yintai Zhang for many stimulating discussions and valuable comments. This research was financially supported by the National Science Centre of Poland (NCN) within the Sonata Bis program (grant No. 2019/34/E/ST2/00440).  M.K. acknowledges support from the NCN under the Maestro Grant No. DEC-2019/34/A/ST2/00081.
\end{acknowledgments}

\section*{Author Contributions}
Y.S., M.K., A.D.F., and S.P. conceived the experiment. Y.S., M.K., and A.D.F. performed the experiments and collected the experimental data. Y.S. and M.K. developed the theoretical framework and numerical modeling. Y.S. analyzed the data and performed the numerical simulations. Y.S. and M.K. prepared the figures and wrote the initial draft of the manuscript. Y.S., M.K.,  A.D.F., and S.P. contributed to the interpretation of the results and revised the manuscript. S.P. supervised the project and acquired funding. All authors discussed the results and contributed to the final manuscript.

\section*{Competing Interests}
The authors declare no competing interests.

\bibliography{bibliography.bib}

\begin{thebibliography}{8}%
\makeatletter
\providecommand \@ifxundefined [1]{%
 \@ifx{#1\undefined}
}%
\providecommand \@ifnum [1]{%
 \ifnum #1\expandafter \@firstoftwo
 \else \expandafter \@secondoftwo
 \fi
}%
\providecommand \@ifx [1]{%
 \ifx #1\expandafter \@firstoftwo
 \else \expandafter \@secondoftwo
 \fi
}%
\providecommand \natexlab [1]{#1}%
\providecommand \enquote  [1]{``#1''}%
\providecommand \bibnamefont  [1]{#1}%
\providecommand \bibfnamefont [1]{#1}%
\providecommand \citenamefont [1]{#1}%
\providecommand \href@noop [0]{\@secondoftwo}%
\providecommand \href [0]{\begingroup \@sanitize@url \@href}%
\providecommand \@href[1]{\@@startlink{#1}\@@href}%
\providecommand \@@href[1]{\endgroup#1\@@endlink}%
\providecommand \@sanitize@url [0]{\catcode `\\12\catcode `\$12\catcode `\&12\catcode `\#12\catcode `\^12\catcode `\_12\catcode `\%12\relax}%
\providecommand \@@startlink[1]{}%
\providecommand \@@endlink[0]{}%
\providecommand \url  [0]{\begingroup\@sanitize@url \@url }%
\providecommand \@url [1]{\endgroup\@href {#1}{\urlprefix }}%
\providecommand \urlprefix  [0]{URL }%
\providecommand \Eprint [0]{\href }%
\providecommand \doibase [0]{https://doi.org/}%
\providecommand \selectlanguage [0]{\@gobble}%
\providecommand \bibinfo  [0]{\@secondoftwo}%
\providecommand \bibfield  [0]{\@secondoftwo}%
\providecommand \translation [1]{[#1]}%
\providecommand \BibitemOpen [0]{}%
\providecommand \bibitemStop [0]{}%
\providecommand \bibitemNoStop [0]{.\EOS\space}%
\providecommand \EOS [0]{\spacefactor3000\relax}%
\providecommand \BibitemShut  [1]{\csname bibitem#1\endcsname}%
\let\auto@bib@innerbib\@empty
\bibitem [{\citenamefont {Bertlmann}\ and\ \citenamefont {Krammer}(2008)}]{bertlmann2008bloch}%
  \BibitemOpen
  \bibfield  {author} {\bibinfo {author} {\bibfnamefont {R.~A.}\ \bibnamefont {Bertlmann}}\ and\ \bibinfo {author} {\bibfnamefont {P.}~\bibnamefont {Krammer}},\ }\bibfield  {title} {\bibinfo {title} {Bloch vectors for qudits},\ }\href@noop {} {\bibfield  {journal} {\bibinfo  {journal} {J. Phys. A: Math. Theor.}\ }\textbf {\bibinfo {volume} {41}},\ \bibinfo {pages} {235303} (\bibinfo {year} {2008})}\BibitemShut {NoStop}%
\bibitem [{\citenamefont {Benenti}\ and\ \citenamefont {Strini}(2009)}]{Benenti2009}%
  \BibitemOpen
  \bibfield  {author} {\bibinfo {author} {\bibfnamefont {G.}~\bibnamefont {Benenti}}\ and\ \bibinfo {author} {\bibfnamefont {G.}~\bibnamefont {Strini}},\ }\bibfield  {title} {\bibinfo {title} {Simple representation of quantum process tomography},\ }\href {https://doi.org/10.1103/PhysRevA.80.022318} {\bibfield  {journal} {\bibinfo  {journal} {Phys. Rev. A}\ }\textbf {\bibinfo {volume} {80}},\ \bibinfo {pages} {022318} (\bibinfo {year} {2009})}\BibitemShut {NoStop}%
\bibitem [{\citenamefont {Gorini}\ \emph {et~al.}(1976)\citenamefont {Gorini}, \citenamefont {Kossakowski},\ and\ \citenamefont {Sudarshan}}]{gorini1976completely}%
  \BibitemOpen
  \bibfield  {author} {\bibinfo {author} {\bibfnamefont {V.}~\bibnamefont {Gorini}}, \bibinfo {author} {\bibfnamefont {A.}~\bibnamefont {Kossakowski}},\ and\ \bibinfo {author} {\bibfnamefont {E.~C.~G.}\ \bibnamefont {Sudarshan}},\ }\bibfield  {title} {\bibinfo {title} {Completely positive dynamical semigroups of n-level systems},\ }\href@noop {} {\bibfield  {journal} {\bibinfo  {journal} {J. Math. Phys.}\ }\textbf {\bibinfo {volume} {17}},\ \bibinfo {pages} {821} (\bibinfo {year} {1976})}\BibitemShut {NoStop}%
\bibitem [{\citenamefont {Kopciuch}\ \emph {et~al.}(2024)\citenamefont {Kopciuch}, \citenamefont {Smolis}, \citenamefont {Miranowicz},\ and\ \citenamefont {Pustelny}}]{Kopciuch2024}%
  \BibitemOpen
  \bibfield  {author} {\bibinfo {author} {\bibfnamefont {M.}~\bibnamefont {Kopciuch}}, \bibinfo {author} {\bibfnamefont {M.}~\bibnamefont {Smolis}}, \bibinfo {author} {\bibfnamefont {A.}~\bibnamefont {Miranowicz}},\ and\ \bibinfo {author} {\bibfnamefont {S.}~\bibnamefont {Pustelny}},\ }\bibfield  {title} {\bibinfo {title} {Optimized optical tomography of quantum states of a room-temperature alkali-metal vapor},\ }\href@noop {} {\bibfield  {journal} {\bibinfo  {journal} {Phys. Rev. A}\ }\textbf {\bibinfo {volume} {109}},\ \bibinfo {pages} {032402} (\bibinfo {year} {2024})}\BibitemShut {NoStop}%
\bibitem [{\citenamefont {Choi}(1975)}]{choi1975completely}%
  \BibitemOpen
  \bibfield  {author} {\bibinfo {author} {\bibfnamefont {M.-D.}\ \bibnamefont {Choi}},\ }\bibfield  {title} {\bibinfo {title} {Completely positive linear maps on complex matrices},\ }\href@noop {} {\bibfield  {journal} {\bibinfo  {journal} {Linear algebra and its applications}\ }\textbf {\bibinfo {volume} {10}},\ \bibinfo {pages} {285} (\bibinfo {year} {1975})}\BibitemShut {NoStop}%
\bibitem [{\citenamefont {Jamio{\l}kowski}(1972)}]{jamiolkowski1972linear}%
  \BibitemOpen
  \bibfield  {author} {\bibinfo {author} {\bibfnamefont {A.}~\bibnamefont {Jamio{\l}kowski}},\ }\bibfield  {title} {\bibinfo {title} {Linear transformations which preserve trace and positive semidefiniteness of operators},\ }\href@noop {} {\bibfield  {journal} {\bibinfo  {journal} {Rep. Math. Phys.}\ }\textbf {\bibinfo {volume} {3}},\ \bibinfo {pages} {275} (\bibinfo {year} {1972})}\BibitemShut {NoStop}%
\bibitem [{\citenamefont {Schmied}(2016)}]{schmied2016quantum}%
  \BibitemOpen
  \bibfield  {author} {\bibinfo {author} {\bibfnamefont {R.}~\bibnamefont {Schmied}},\ }\bibfield  {title} {\bibinfo {title} {Quantum state tomography of a single qubit: comparison of methods},\ }\href@noop {} {\bibfield  {journal} {\bibinfo  {journal} {J. Mod. Opt.}\ }\textbf {\bibinfo {volume} {63}},\ \bibinfo {pages} {1744} (\bibinfo {year} {2016})}\BibitemShut {NoStop}%
\bibitem [{\citenamefont {Schwemmer}\ \emph {et~al.}(2015)\citenamefont {Schwemmer}, \citenamefont {Knips}, \citenamefont {Richart}, \citenamefont {Weinfurter}, \citenamefont {Moroder}, \citenamefont {Kleinmann},\ and\ \citenamefont {G{\"u}hne}}]{schwemmer2015systematic}%
  \BibitemOpen
  \bibfield  {author} {\bibinfo {author} {\bibfnamefont {C.}~\bibnamefont {Schwemmer}}, \bibinfo {author} {\bibfnamefont {L.}~\bibnamefont {Knips}}, \bibinfo {author} {\bibfnamefont {D.}~\bibnamefont {Richart}}, \bibinfo {author} {\bibfnamefont {H.}~\bibnamefont {Weinfurter}}, \bibinfo {author} {\bibfnamefont {T.}~\bibnamefont {Moroder}}, \bibinfo {author} {\bibfnamefont {M.}~\bibnamefont {Kleinmann}},\ and\ \bibinfo {author} {\bibfnamefont {O.}~\bibnamefont {G{\"u}hne}},\ }\bibfield  {title} {\bibinfo {title} {Systematic errors in current quantum state tomography tools},\ }\href@noop {} {\bibfield  {journal} {\bibinfo  {journal} {Phys. Rev. Lett.}\ }\textbf {\bibinfo {volume} {114}},\ \bibinfo {pages} {080403} (\bibinfo {year} {2015})}\BibitemShut {NoStop}%
\end{thebibliography}%


\begin{thebibliography}{57}%
\makeatletter
\providecommand \@ifxundefined [1]{%
 \@ifx{#1\undefined}
}%
\providecommand \@ifnum [1]{%
 \ifnum #1\expandafter \@firstoftwo
 \else \expandafter \@secondoftwo
 \fi
}%
\providecommand \@ifx [1]{%
 \ifx #1\expandafter \@firstoftwo
 \else \expandafter \@secondoftwo
 \fi
}%
\providecommand \natexlab [1]{#1}%
\providecommand \enquote  [1]{``#1''}%
\providecommand \bibnamefont  [1]{#1}%
\providecommand \bibfnamefont [1]{#1}%
\providecommand \citenamefont [1]{#1}%
\providecommand \href@noop [0]{\@secondoftwo}%
\providecommand \href [0]{\begingroup \@sanitize@url \@href}%
\providecommand \@href[1]{\@@startlink{#1}\@@href}%
\providecommand \@@href[1]{\endgroup#1\@@endlink}%
\providecommand \@sanitize@url [0]{\catcode `\\12\catcode `\$12\catcode `\&12\catcode `\#12\catcode `\^12\catcode `\_12\catcode `\%12\relax}%
\providecommand \@@startlink[1]{}%
\providecommand \@@endlink[0]{}%
\providecommand \url  [0]{\begingroup\@sanitize@url \@url }%
\providecommand \@url [1]{\endgroup\@href {#1}{\urlprefix }}%
\providecommand \urlprefix  [0]{URL }%
\providecommand \Eprint [0]{\href }%
\providecommand \doibase [0]{https://doi.org/}%
\providecommand \selectlanguage [0]{\@gobble}%
\providecommand \bibinfo  [0]{\@secondoftwo}%
\providecommand \bibfield  [0]{\@secondoftwo}%
\providecommand \translation [1]{[#1]}%
\providecommand \BibitemOpen [0]{}%
\providecommand \bibitemStop [0]{}%
\providecommand \bibitemNoStop [0]{.\EOS\space}%
\providecommand \EOS [0]{\spacefactor3000\relax}%
\providecommand \BibitemShut  [1]{\csname bibitem#1\endcsname}%
\let\auto@bib@innerbib\@empty
\bibitem [{\citenamefont {Degen}\ \emph {et~al.}(2017)\citenamefont {Degen}, \citenamefont {Reinhard},\ and\ \citenamefont {Cappellaro}}]{degen2017quantum}%
  \BibitemOpen
  \bibfield  {author} {\bibinfo {author} {\bibfnamefont {C.~L.}\ \bibnamefont {Degen}}, \bibinfo {author} {\bibfnamefont {F.}~\bibnamefont {Reinhard}},\ and\ \bibinfo {author} {\bibfnamefont {P.}~\bibnamefont {Cappellaro}},\ }\bibfield  {title} {\bibinfo {title} {Quantum sensing},\ }\href@noop {} {\bibfield  {journal} {\bibinfo  {journal} {Rev. Mod. Phys.}\ }\textbf {\bibinfo {volume} {89}},\ \bibinfo {pages} {035002} (\bibinfo {year} {2017})}\BibitemShut {NoStop}%
\bibitem [{\citenamefont {Giovannetti}\ \emph {et~al.}(2011)\citenamefont {Giovannetti}, \citenamefont {Lloyd},\ and\ \citenamefont {Maccone}}]{giovannetti2011advances}%
  \BibitemOpen
  \bibfield  {author} {\bibinfo {author} {\bibfnamefont {V.}~\bibnamefont {Giovannetti}}, \bibinfo {author} {\bibfnamefont {S.}~\bibnamefont {Lloyd}},\ and\ \bibinfo {author} {\bibfnamefont {L.}~\bibnamefont {Maccone}},\ }\bibfield  {title} {\bibinfo {title} {Advances in quantum metrology},\ }\href@noop {} {\bibfield  {journal} {\bibinfo  {journal} {Nat. Photonics}\ }\textbf {\bibinfo {volume} {5}},\ \bibinfo {pages} {222} (\bibinfo {year} {2011})}\BibitemShut {NoStop}%
\bibitem [{\citenamefont {Nielsen}\ and\ \citenamefont {Chuang}(2010)}]{nielsen2010quantum}%
  \BibitemOpen
  \bibfield  {author} {\bibinfo {author} {\bibfnamefont {M.~A.}\ \bibnamefont {Nielsen}}\ and\ \bibinfo {author} {\bibfnamefont {I.~L.}\ \bibnamefont {Chuang}},\ }\href@noop {} {\emph {\bibinfo {title} {Quantum computation and quantum information}}}\ (\bibinfo  {publisher} {Cambridge University Press},\ \bibinfo {year} {2010})\BibitemShut {NoStop}%
\bibitem [{\citenamefont {Lobino}\ \emph {et~al.}(2008)\citenamefont {Lobino}, \citenamefont {Korystov}, \citenamefont {Kupchak}, \citenamefont {Figueroa}, \citenamefont {Sanders},\ and\ \citenamefont {Lvovsky}}]{lobino2008complete}%
  \BibitemOpen
  \bibfield  {author} {\bibinfo {author} {\bibfnamefont {M.}~\bibnamefont {Lobino}}, \bibinfo {author} {\bibfnamefont {D.}~\bibnamefont {Korystov}}, \bibinfo {author} {\bibfnamefont {C.}~\bibnamefont {Kupchak}}, \bibinfo {author} {\bibfnamefont {E.}~\bibnamefont {Figueroa}}, \bibinfo {author} {\bibfnamefont {B.~C.}\ \bibnamefont {Sanders}},\ and\ \bibinfo {author} {\bibfnamefont {A.}~\bibnamefont {Lvovsky}},\ }\bibfield  {title} {\bibinfo {title} {Complete characterization of quantum-optical processes},\ }\href@noop {} {\bibfield  {journal} {\bibinfo  {journal} {Science}\ }\textbf {\bibinfo {volume} {322}},\ \bibinfo {pages} {563} (\bibinfo {year} {2008})}\BibitemShut {NoStop}%
\bibitem [{\citenamefont {Mohseni}\ \emph {et~al.}(2008)\citenamefont {Mohseni}, \citenamefont {Rezakhani},\ and\ \citenamefont {Lidar}}]{mohseni2008quantum}%
  \BibitemOpen
  \bibfield  {author} {\bibinfo {author} {\bibfnamefont {M.}~\bibnamefont {Mohseni}}, \bibinfo {author} {\bibfnamefont {A.~T.}\ \bibnamefont {Rezakhani}},\ and\ \bibinfo {author} {\bibfnamefont {D.~A.}\ \bibnamefont {Lidar}},\ }\bibfield  {title} {\bibinfo {title} {Quantum-process tomography: Resource analysis of different strategies},\ }\href@noop {} {\bibfield  {journal} {\bibinfo  {journal} {Phys. Rev. A}\ }\textbf {\bibinfo {volume} {77}},\ \bibinfo {pages} {032322} (\bibinfo {year} {2008})}\BibitemShut {NoStop}%
\bibitem [{\citenamefont {Keesling}\ \emph {et~al.}(2019)\citenamefont {Keesling} \emph {et~al.}}]{keesling2019quantum}%
  \BibitemOpen
  \bibfield  {author} {\bibinfo {author} {\bibfnamefont {A.}~\bibnamefont {Keesling}} \emph {et~al.},\ }\bibfield  {title} {\bibinfo {title} {Quantum {K}ibble--{Z}urek mechanism and critical dynamics on a programmable {Rydberg} simulator},\ }\href@noop {} {\bibfield  {journal} {\bibinfo  {journal} {Nature (London)}\ }\textbf {\bibinfo {volume} {568}},\ \bibinfo {pages} {207} (\bibinfo {year} {2019})}\BibitemShut {NoStop}%
\bibitem [{\citenamefont {Ebadi}\ \emph {et~al.}(2021)\citenamefont {Ebadi} \emph {et~al.}}]{ebadi2021quantum}%
  \BibitemOpen
  \bibfield  {author} {\bibinfo {author} {\bibfnamefont {S.}~\bibnamefont {Ebadi}} \emph {et~al.},\ }\bibfield  {title} {\bibinfo {title} {Quantum phases of matter on a 256-atom programmable quantum simulator},\ }\href@noop {} {\bibfield  {journal} {\bibinfo  {journal} {Nature (London)}\ }\textbf {\bibinfo {volume} {595}},\ \bibinfo {pages} {227} (\bibinfo {year} {2021})}\BibitemShut {NoStop}%
\bibitem [{\citenamefont {Chaudhury}\ \emph {et~al.}(2007)\citenamefont {Chaudhury}, \citenamefont {Merkel}, \citenamefont {Herr}, \citenamefont {Silberfarb}, \citenamefont {Deutsch},\ and\ \citenamefont {Jessen}}]{chaudhury2007quantum}%
  \BibitemOpen
  \bibfield  {author} {\bibinfo {author} {\bibfnamefont {S.}~\bibnamefont {Chaudhury}}, \bibinfo {author} {\bibfnamefont {S.}~\bibnamefont {Merkel}}, \bibinfo {author} {\bibfnamefont {T.}~\bibnamefont {Herr}}, \bibinfo {author} {\bibfnamefont {A.}~\bibnamefont {Silberfarb}}, \bibinfo {author} {\bibfnamefont {I.~H.}\ \bibnamefont {Deutsch}},\ and\ \bibinfo {author} {\bibfnamefont {P.~S.}\ \bibnamefont {Jessen}},\ }\bibfield  {title} {\bibinfo {title} {Quantum control of the hyperfine spin of a {Cs} atom ensemble},\ }\href@noop {} {\bibfield  {journal} {\bibinfo  {journal} {Phys. Rev. Lett.}\ }\textbf {\bibinfo {volume} {99}},\ \bibinfo {pages} {163002} (\bibinfo {year} {2007})}\BibitemShut {NoStop}%
\bibitem [{\citenamefont {Saffman}\ \emph {et~al.}(2010)\citenamefont {Saffman}, \citenamefont {Walker},\ and\ \citenamefont {M{\o}lmer}}]{saffman2010quantum}%
  \BibitemOpen
  \bibfield  {author} {\bibinfo {author} {\bibfnamefont {M.}~\bibnamefont {Saffman}}, \bibinfo {author} {\bibfnamefont {T.~G.}\ \bibnamefont {Walker}},\ and\ \bibinfo {author} {\bibfnamefont {K.}~\bibnamefont {M{\o}lmer}},\ }\bibfield  {title} {\bibinfo {title} {Quantum information with {R}ydberg atoms},\ }\href@noop {} {\bibfield  {journal} {\bibinfo  {journal} {Rev. Mod. Phys.}\ }\textbf {\bibinfo {volume} {82}},\ \bibinfo {pages} {2313} (\bibinfo {year} {2010})}\BibitemShut {NoStop}%
\bibitem [{\citenamefont {Ben~Av}\ \emph {et~al.}(2020)\citenamefont {Ben~Av}, \citenamefont {Shapira}, \citenamefont {Akerman},\ and\ \citenamefont {Ozeri}}]{ben2020direct}%
  \BibitemOpen
  \bibfield  {author} {\bibinfo {author} {\bibfnamefont {E.}~\bibnamefont {Ben~Av}}, \bibinfo {author} {\bibfnamefont {Y.}~\bibnamefont {Shapira}}, \bibinfo {author} {\bibfnamefont {N.}~\bibnamefont {Akerman}},\ and\ \bibinfo {author} {\bibfnamefont {R.}~\bibnamefont {Ozeri}},\ }\bibfield  {title} {\bibinfo {title} {Direct reconstruction of the quantum-master-equation dynamics of a trapped-ion qubit},\ }\href@noop {} {\bibfield  {journal} {\bibinfo  {journal} {Phys. Rev. A}\ }\textbf {\bibinfo {volume} {101}},\ \bibinfo {pages} {062305} (\bibinfo {year} {2020})}\BibitemShut {NoStop}%
\bibitem [{\citenamefont {Figgatt}\ \emph {et~al.}(2019)\citenamefont {Figgatt}, \citenamefont {Ostrander}, \citenamefont {Linke}, \citenamefont {Landsman}, \citenamefont {Zhu}, \citenamefont {Maslov},\ and\ \citenamefont {Monroe}}]{figgatt2019parallel}%
  \BibitemOpen
  \bibfield  {author} {\bibinfo {author} {\bibfnamefont {C.}~\bibnamefont {Figgatt}}, \bibinfo {author} {\bibfnamefont {A.}~\bibnamefont {Ostrander}}, \bibinfo {author} {\bibfnamefont {N.~M.}\ \bibnamefont {Linke}}, \bibinfo {author} {\bibfnamefont {K.~A.}\ \bibnamefont {Landsman}}, \bibinfo {author} {\bibfnamefont {D.}~\bibnamefont {Zhu}}, \bibinfo {author} {\bibfnamefont {D.}~\bibnamefont {Maslov}},\ and\ \bibinfo {author} {\bibfnamefont {C.}~\bibnamefont {Monroe}},\ }\bibfield  {title} {\bibinfo {title} {Parallel entangling operations on a universal ion-trap quantum computer},\ }\href@noop {} {\bibfield  {journal} {\bibinfo  {journal} {Nature (London)}\ }\textbf {\bibinfo {volume} {572}},\ \bibinfo {pages} {368} (\bibinfo {year} {2019})}\BibitemShut {NoStop}%
\bibitem [{\citenamefont {Nam}\ \emph {et~al.}(2020)\citenamefont {Nam} \emph {et~al.}}]{nam2020ground}%
  \BibitemOpen
  \bibfield  {author} {\bibinfo {author} {\bibfnamefont {Y.}~\bibnamefont {Nam}} \emph {et~al.},\ }\bibfield  {title} {\bibinfo {title} {Ground-state energy estimation of the water molecule on a trapped-ion quantum computer},\ }\href@noop {} {\bibfield  {journal} {\bibinfo  {journal} {npj Quantum Inf.}\ }\textbf {\bibinfo {volume} {6}},\ \bibinfo {pages} {33} (\bibinfo {year} {2020})}\BibitemShut {NoStop}%
\bibitem [{\citenamefont {Howard}\ \emph {et~al.}(2006)\citenamefont {Howard}, \citenamefont {Twamley}, \citenamefont {Wittmann}, \citenamefont {Gaebel}, \citenamefont {Jelezko},\ and\ \citenamefont {Wrachtrup}}]{howard2006quantum}%
  \BibitemOpen
  \bibfield  {author} {\bibinfo {author} {\bibfnamefont {M.}~\bibnamefont {Howard}}, \bibinfo {author} {\bibfnamefont {J.}~\bibnamefont {Twamley}}, \bibinfo {author} {\bibfnamefont {C.}~\bibnamefont {Wittmann}}, \bibinfo {author} {\bibfnamefont {T.}~\bibnamefont {Gaebel}}, \bibinfo {author} {\bibfnamefont {F.}~\bibnamefont {Jelezko}},\ and\ \bibinfo {author} {\bibfnamefont {J.}~\bibnamefont {Wrachtrup}},\ }\bibfield  {title} {\bibinfo {title} {Quantum process tomography and {L}inblad estimation of a solid-state qubit},\ }\href@noop {} {\bibfield  {journal} {\bibinfo  {journal} {New J. Phys.}\ }\textbf {\bibinfo {volume} {8}},\ \bibinfo {pages} {33} (\bibinfo {year} {2006})}\BibitemShut {NoStop}%
\bibitem [{\citenamefont {Zhou}\ \emph {et~al.}(2015)\citenamefont {Zhou}, \citenamefont {Cable}, \citenamefont {Whittaker}, \citenamefont {Shadbolt}, \citenamefont {O’Brien},\ and\ \citenamefont {Matthews}}]{zhou2015quantum}%
  \BibitemOpen
  \bibfield  {author} {\bibinfo {author} {\bibfnamefont {X.-Q.}\ \bibnamefont {Zhou}}, \bibinfo {author} {\bibfnamefont {H.}~\bibnamefont {Cable}}, \bibinfo {author} {\bibfnamefont {R.}~\bibnamefont {Whittaker}}, \bibinfo {author} {\bibfnamefont {P.}~\bibnamefont {Shadbolt}}, \bibinfo {author} {\bibfnamefont {J.~L.}\ \bibnamefont {O’Brien}},\ and\ \bibinfo {author} {\bibfnamefont {J.~C.}\ \bibnamefont {Matthews}},\ }\bibfield  {title} {\bibinfo {title} {Quantum-enhanced tomography of unitary processes},\ }\href@noop {} {\bibfield  {journal} {\bibinfo  {journal} {Optica}\ }\textbf {\bibinfo {volume} {2}},\ \bibinfo {pages} {510} (\bibinfo {year} {2015})}\BibitemShut {NoStop}%
\bibitem [{\citenamefont {Wang}\ \emph {et~al.}(2023)\citenamefont {Wang}, \citenamefont {Zhan}, \citenamefont {Li}, \citenamefont {Xiao}, \citenamefont {Zhu}, \citenamefont {Qu}, \citenamefont {Lin}, \citenamefont {Yu},\ and\ \citenamefont {Xue}}]{wang2023generalized}%
  \BibitemOpen
  \bibfield  {author} {\bibinfo {author} {\bibfnamefont {X.}~\bibnamefont {Wang}}, \bibinfo {author} {\bibfnamefont {X.}~\bibnamefont {Zhan}}, \bibinfo {author} {\bibfnamefont {Y.}~\bibnamefont {Li}}, \bibinfo {author} {\bibfnamefont {L.}~\bibnamefont {Xiao}}, \bibinfo {author} {\bibfnamefont {G.}~\bibnamefont {Zhu}}, \bibinfo {author} {\bibfnamefont {D.}~\bibnamefont {Qu}}, \bibinfo {author} {\bibfnamefont {Q.}~\bibnamefont {Lin}}, \bibinfo {author} {\bibfnamefont {Y.}~\bibnamefont {Yu}},\ and\ \bibinfo {author} {\bibfnamefont {P.}~\bibnamefont {Xue}},\ }\bibfield  {title} {\bibinfo {title} {Generalized quantum measurements on a higher-dimensional system via quantum walks},\ }\href@noop {} {\bibfield  {journal} {\bibinfo  {journal} {Phys. Rev. Lett.}\ }\textbf {\bibinfo {volume} {131}},\ \bibinfo {pages} {150803} (\bibinfo {year} {2023})}\BibitemShut {NoStop}%
\bibitem [{\citenamefont {Wang}\ \emph {et~al.}(2024)\citenamefont {Wang}, \citenamefont {Lyu}, \citenamefont {Liu},\ and\ \citenamefont {Wang}}]{wang2024polarization}%
  \BibitemOpen
  \bibfield  {author} {\bibinfo {author} {\bibfnamefont {Q.}~\bibnamefont {Wang}}, \bibinfo {author} {\bibfnamefont {D.}~\bibnamefont {Lyu}}, \bibinfo {author} {\bibfnamefont {J.}~\bibnamefont {Liu}},\ and\ \bibinfo {author} {\bibfnamefont {J.}~\bibnamefont {Wang}},\ }\bibfield  {title} {\bibinfo {title} {Polarization and orbital angular momentum encoded quantum toffoli gate enabled by diffractive neural networks},\ }\href@noop {} {\bibfield  {journal} {\bibinfo  {journal} {Phys. Rev. Lett.}\ }\textbf {\bibinfo {volume} {133}},\ \bibinfo {pages} {140601} (\bibinfo {year} {2024})}\BibitemShut {NoStop}%
\bibitem [{\citenamefont {Arute}\ \emph {et~al.}(2019)\citenamefont {Arute} \emph {et~al.}}]{arute2019quantum}%
  \BibitemOpen
  \bibfield  {author} {\bibinfo {author} {\bibfnamefont {F.}~\bibnamefont {Arute}} \emph {et~al.},\ }\bibfield  {title} {\bibinfo {title} {Quantum supremacy using a programmable superconducting processor},\ }\href@noop {} {\bibfield  {journal} {\bibinfo  {journal} {Nature (London)}\ }\textbf {\bibinfo {volume} {574}},\ \bibinfo {pages} {505} (\bibinfo {year} {2019})}\BibitemShut {NoStop}%
\bibitem [{\citenamefont {Samach}\ \emph {et~al.}(2022)\citenamefont {Samach} \emph {et~al.}}]{samach2022lindblad}%
  \BibitemOpen
  \bibfield  {author} {\bibinfo {author} {\bibfnamefont {G.~O.}\ \bibnamefont {Samach}} \emph {et~al.},\ }\bibfield  {title} {\bibinfo {title} {Lindblad tomography of a superconducting quantum processor},\ }\href@noop {} {\bibfield  {journal} {\bibinfo  {journal} {Phys. Rev. Appl.}\ }\textbf {\bibinfo {volume} {18}},\ \bibinfo {pages} {064056} (\bibinfo {year} {2022})}\BibitemShut {NoStop}%
\bibitem [{\citenamefont {Thew}\ \emph {et~al.}(2002)\citenamefont {Thew}, \citenamefont {Nemoto}, \citenamefont {White},\ and\ \citenamefont {Munro}}]{thew2002qudit}%
  \BibitemOpen
  \bibfield  {author} {\bibinfo {author} {\bibfnamefont {R.~T.}\ \bibnamefont {Thew}}, \bibinfo {author} {\bibfnamefont {K.}~\bibnamefont {Nemoto}}, \bibinfo {author} {\bibfnamefont {A.~G.}\ \bibnamefont {White}},\ and\ \bibinfo {author} {\bibfnamefont {W.~J.}\ \bibnamefont {Munro}},\ }\bibfield  {title} {\bibinfo {title} {Qudit quantum-state tomography},\ }\href@noop {} {\bibfield  {journal} {\bibinfo  {journal} {Phys. Rev. A}\ }\textbf {\bibinfo {volume} {66}},\ \bibinfo {pages} {012303} (\bibinfo {year} {2002})}\BibitemShut {NoStop}%
\bibitem [{\citenamefont {Wang}\ \emph {et~al.}(2020)\citenamefont {Wang}, \citenamefont {Hu}, \citenamefont {Sanders},\ and\ \citenamefont {Kais}}]{wang2020qudits}%
  \BibitemOpen
  \bibfield  {author} {\bibinfo {author} {\bibfnamefont {Y.}~\bibnamefont {Wang}}, \bibinfo {author} {\bibfnamefont {Z.}~\bibnamefont {Hu}}, \bibinfo {author} {\bibfnamefont {B.~C.}\ \bibnamefont {Sanders}},\ and\ \bibinfo {author} {\bibfnamefont {S.}~\bibnamefont {Kais}},\ }\bibfield  {title} {\bibinfo {title} {Qudits and high-dimensional quantum computing},\ }\href@noop {} {\bibfield  {journal} {\bibinfo  {journal} {Front. Phys.}\ }\textbf {\bibinfo {volume} {8}},\ \bibinfo {pages} {589504} (\bibinfo {year} {2020})}\BibitemShut {NoStop}%
\bibitem [{\citenamefont {Klimov}\ \emph {et~al.}(2003)\citenamefont {Klimov}, \citenamefont {Guzm{\'a}n}, \citenamefont {Retamal},\ and\ \citenamefont {Saavedra}}]{klimov2003qutrit}%
  \BibitemOpen
  \bibfield  {author} {\bibinfo {author} {\bibfnamefont {A.}~\bibnamefont {Klimov}}, \bibinfo {author} {\bibfnamefont {R.}~\bibnamefont {Guzm{\'a}n}}, \bibinfo {author} {\bibfnamefont {J.}~\bibnamefont {Retamal}},\ and\ \bibinfo {author} {\bibfnamefont {C.}~\bibnamefont {Saavedra}},\ }\bibfield  {title} {\bibinfo {title} {Qutrit quantum computer with trapped ions},\ }\href@noop {} {\bibfield  {journal} {\bibinfo  {journal} {Phys. Rev. A}\ }\textbf {\bibinfo {volume} {67}},\ \bibinfo {pages} {062313} (\bibinfo {year} {2003})}\BibitemShut {NoStop}%
\bibitem [{\citenamefont {D'Ariano}\ and\ \citenamefont {Presti}(2001)}]{d2001quantum}%
  \BibitemOpen
  \bibfield  {author} {\bibinfo {author} {\bibfnamefont {G.}~\bibnamefont {D'Ariano}}\ and\ \bibinfo {author} {\bibfnamefont {P.~L.}\ \bibnamefont {Presti}},\ }\bibfield  {title} {\bibinfo {title} {Quantum tomography for measuring experimentally the matrix elements of an arbitrary quantum operation},\ }\href@noop {} {\bibfield  {journal} {\bibinfo  {journal} {Phys. Rev. Lett.}\ }\textbf {\bibinfo {volume} {86}},\ \bibinfo {pages} {4195} (\bibinfo {year} {2001})}\BibitemShut {NoStop}%
\bibitem [{\citenamefont {White}\ \emph {et~al.}(2022)\citenamefont {White}, \citenamefont {Pollock}, \citenamefont {Hollenberg}, \citenamefont {Modi},\ and\ \citenamefont {Hill}}]{white2022non}%
  \BibitemOpen
  \bibfield  {author} {\bibinfo {author} {\bibfnamefont {G.~A.}\ \bibnamefont {White}}, \bibinfo {author} {\bibfnamefont {F.~A.}\ \bibnamefont {Pollock}}, \bibinfo {author} {\bibfnamefont {L.~C.}\ \bibnamefont {Hollenberg}}, \bibinfo {author} {\bibfnamefont {K.}~\bibnamefont {Modi}},\ and\ \bibinfo {author} {\bibfnamefont {C.~D.}\ \bibnamefont {Hill}},\ }\bibfield  {title} {\bibinfo {title} {Non-{M}arkovian quantum process tomography},\ }\href@noop {} {\bibfield  {journal} {\bibinfo  {journal} {PRX Quantum}\ }\textbf {\bibinfo {volume} {3}},\ \bibinfo {pages} {020344} (\bibinfo {year} {2022})}\BibitemShut {NoStop}%
\bibitem [{\citenamefont {Ahmed}\ \emph {et~al.}(2023)\citenamefont {Ahmed}, \citenamefont {Quijandr{\'\i}a},\ and\ \citenamefont {Kockum}}]{ahmed2023gradient}%
  \BibitemOpen
  \bibfield  {author} {\bibinfo {author} {\bibfnamefont {S.}~\bibnamefont {Ahmed}}, \bibinfo {author} {\bibfnamefont {F.}~\bibnamefont {Quijandr{\'\i}a}},\ and\ \bibinfo {author} {\bibfnamefont {A.~F.}\ \bibnamefont {Kockum}},\ }\bibfield  {title} {\bibinfo {title} {Gradient-descent quantum process tomography by learning kraus operators},\ }\href@noop {} {\bibfield  {journal} {\bibinfo  {journal} {Phys. Rev. Lett.}\ }\textbf {\bibinfo {volume} {130}},\ \bibinfo {pages} {150402} (\bibinfo {year} {2023})}\BibitemShut {NoStop}%
\bibitem [{\citenamefont {Di~Colandrea}\ \emph {et~al.}(2024)\citenamefont {Di~Colandrea}, \citenamefont {Dehghan}, \citenamefont {D’Errico},\ and\ \citenamefont {Karimi}}]{di2024fourier}%
  \BibitemOpen
  \bibfield  {author} {\bibinfo {author} {\bibfnamefont {F.}~\bibnamefont {Di~Colandrea}}, \bibinfo {author} {\bibfnamefont {N.}~\bibnamefont {Dehghan}}, \bibinfo {author} {\bibfnamefont {A.}~\bibnamefont {D’Errico}},\ and\ \bibinfo {author} {\bibfnamefont {E.}~\bibnamefont {Karimi}},\ }\bibfield  {title} {\bibinfo {title} {Fourier quantum process tomography},\ }\href@noop {} {\bibfield  {journal} {\bibinfo  {journal} {npj Quantum Inf.}\ }\textbf {\bibinfo {volume} {10}},\ \bibinfo {pages} {49} (\bibinfo {year} {2024})}\BibitemShut {NoStop}%
\bibitem [{\citenamefont {Bu{\v{z}}ek}(1998)}]{buvzek1998reconstruction}%
  \BibitemOpen
  \bibfield  {author} {\bibinfo {author} {\bibfnamefont {V.}~\bibnamefont {Bu{\v{z}}ek}},\ }\bibfield  {title} {\bibinfo {title} {Reconstruction of {L}iouvillian superoperators},\ }\href@noop {} {\bibfield  {journal} {\bibinfo  {journal} {Phys. Rev. A}\ }\textbf {\bibinfo {volume} {58}},\ \bibinfo {pages} {1723} (\bibinfo {year} {1998})}\BibitemShut {NoStop}%
\bibitem [{\citenamefont {Hayden}\ and\ \citenamefont {Sorce}(2022)}]{hayden2022canonical}%
  \BibitemOpen
  \bibfield  {author} {\bibinfo {author} {\bibfnamefont {P.}~\bibnamefont {Hayden}}\ and\ \bibinfo {author} {\bibfnamefont {J.}~\bibnamefont {Sorce}},\ }\bibfield  {title} {\bibinfo {title} {A canonical {H}amiltonian for open quantum systems},\ }\href@noop {} {\bibfield  {journal} {\bibinfo  {journal} {J. Phys. A: Math. Theor.}\ }\textbf {\bibinfo {volume} {55}},\ \bibinfo {pages} {225302} (\bibinfo {year} {2022})}\BibitemShut {NoStop}%
\bibitem [{\citenamefont {Finkelstein}\ \emph {et~al.}(2023)\citenamefont {Finkelstein}, \citenamefont {Bali}, \citenamefont {Firstenberg},\ and\ \citenamefont {Novikova}}]{finkelstein2023practical}%
  \BibitemOpen
  \bibfield  {author} {\bibinfo {author} {\bibfnamefont {R.}~\bibnamefont {Finkelstein}}, \bibinfo {author} {\bibfnamefont {S.}~\bibnamefont {Bali}}, \bibinfo {author} {\bibfnamefont {O.}~\bibnamefont {Firstenberg}},\ and\ \bibinfo {author} {\bibfnamefont {I.}~\bibnamefont {Novikova}},\ }\bibfield  {title} {\bibinfo {title} {A practical guide to electromagnetically induced transparency in atomic vapor},\ }\href@noop {} {\bibfield  {journal} {\bibinfo  {journal} {New J. Phys.}\ }\textbf {\bibinfo {volume} {25}},\ \bibinfo {pages} {035001} (\bibinfo {year} {2023})}\BibitemShut {NoStop}%
\bibitem [{\citenamefont {Li}\ \emph {et~al.}(2022)\citenamefont {Li}, \citenamefont {Dai}, \citenamefont {Liu}, \citenamefont {Xu},\ and\ \citenamefont {Chida}}]{li2022spin}%
  \BibitemOpen
  \bibfield  {author} {\bibinfo {author} {\bibfnamefont {S.}~\bibnamefont {Li}}, \bibinfo {author} {\bibfnamefont {P.}~\bibnamefont {Dai}}, \bibinfo {author} {\bibfnamefont {J.}~\bibnamefont {Liu}}, \bibinfo {author} {\bibfnamefont {Z.}~\bibnamefont {Xu}},\ and\ \bibinfo {author} {\bibfnamefont {K.}~\bibnamefont {Chida}},\ }\bibfield  {title} {\bibinfo {title} {Spin relaxation of rubidium atoms in an octadecyltrichlorosilane anti-relaxation and anti-reflection coated vacuum multipass cell},\ }\href@noop {} {\bibfield  {journal} {\bibinfo  {journal} {Opt. Mater. Express}\ }\textbf {\bibinfo {volume} {12}},\ \bibinfo {pages} {4384} (\bibinfo {year} {2022})}\BibitemShut {NoStop}%
\bibitem [{\citenamefont {Auzinsh}\ \emph {et~al.}(2010)\citenamefont {Auzinsh}, \citenamefont {Budker},\ and\ \citenamefont {Rochester}}]{auzinsh2010optically}%
  \BibitemOpen
  \bibfield  {author} {\bibinfo {author} {\bibfnamefont {M.}~\bibnamefont {Auzinsh}}, \bibinfo {author} {\bibfnamefont {D.}~\bibnamefont {Budker}},\ and\ \bibinfo {author} {\bibfnamefont {S.}~\bibnamefont {Rochester}},\ }\href@noop {} {\emph {\bibinfo {title} {Optically polarized atoms: understanding light-atom interactions}}}\ (\bibinfo  {publisher} {Oxford University Press},\ \bibinfo {year} {2010})\BibitemShut {NoStop}%
\bibitem [{\citenamefont {Lindblad}(1976)}]{lindblad1976generators}%
  \BibitemOpen
  \bibfield  {author} {\bibinfo {author} {\bibfnamefont {G.}~\bibnamefont {Lindblad}},\ }\bibfield  {title} {\bibinfo {title} {On the generators of quantum dynamical semigroups},\ }\href@noop {} {\bibfield  {journal} {\bibinfo  {journal} {Commun. Math. Phys.}\ }\textbf {\bibinfo {volume} {48}},\ \bibinfo {pages} {119} (\bibinfo {year} {1976})}\BibitemShut {NoStop}%
\bibitem [{\citenamefont {Gorini}\ \emph {et~al.}(1976)\citenamefont {Gorini}, \citenamefont {Kossakowski},\ and\ \citenamefont {Sudarshan}}]{gorini1976completely}%
  \BibitemOpen
  \bibfield  {author} {\bibinfo {author} {\bibfnamefont {V.}~\bibnamefont {Gorini}}, \bibinfo {author} {\bibfnamefont {A.}~\bibnamefont {Kossakowski}},\ and\ \bibinfo {author} {\bibfnamefont {E.~C.~G.}\ \bibnamefont {Sudarshan}},\ }\bibfield  {title} {\bibinfo {title} {Completely positive dynamical semigroups of n-level systems},\ }\href@noop {} {\bibfield  {journal} {\bibinfo  {journal} {J. Math. Phys.}\ }\textbf {\bibinfo {volume} {17}},\ \bibinfo {pages} {821} (\bibinfo {year} {1976})}\BibitemShut {NoStop}%
\bibitem [{\citenamefont {Bloch}(1957)}]{bloch1957generalized}%
  \BibitemOpen
  \bibfield  {author} {\bibinfo {author} {\bibfnamefont {F.}~\bibnamefont {Bloch}},\ }\bibfield  {title} {\bibinfo {title} {Generalized theory of relaxation},\ }\href@noop {} {\bibfield  {journal} {\bibinfo  {journal} {Phys. Rev.}\ }\textbf {\bibinfo {volume} {105}},\ \bibinfo {pages} {1206} (\bibinfo {year} {1957})}\BibitemShut {NoStop}%
\bibitem [{\citenamefont {Redfield}(1957)}]{redfield1957theory}%
  \BibitemOpen
  \bibfield  {author} {\bibinfo {author} {\bibfnamefont {A.~G.}\ \bibnamefont {Redfield}},\ }\bibfield  {title} {\bibinfo {title} {On the theory of relaxation processes},\ }\href@noop {} {\bibfield  {journal} {\bibinfo  {journal} {IBM J. Res. Dev.}\ }\textbf {\bibinfo {volume} {1}},\ \bibinfo {pages} {19} (\bibinfo {year} {1957})}\BibitemShut {NoStop}%
\bibitem [{\citenamefont {Davies}(1974)}]{davies1974markovian}%
  \BibitemOpen
  \bibfield  {author} {\bibinfo {author} {\bibfnamefont {E.~B.}\ \bibnamefont {Davies}},\ }\bibfield  {title} {\bibinfo {title} {Markovian master equations},\ }\href@noop {} {\bibfield  {journal} {\bibinfo  {journal} {Commun. Math. Phys.}\ }\textbf {\bibinfo {volume} {39}},\ \bibinfo {pages} {91} (\bibinfo {year} {1974})}\BibitemShut {NoStop}%
\bibitem [{\citenamefont {Singh}\ \emph {et~al.}(2024)\citenamefont {Singh}, \citenamefont {Sawicki},\ and\ \citenamefont {Korbicz}}]{singh2024pointer}%
  \BibitemOpen
  \bibfield  {author} {\bibinfo {author} {\bibfnamefont {U.}~\bibnamefont {Singh}}, \bibinfo {author} {\bibfnamefont {A.}~\bibnamefont {Sawicki}},\ and\ \bibinfo {author} {\bibfnamefont {J.~K.}\ \bibnamefont {Korbicz}},\ }\bibfield  {title} {\bibinfo {title} {Pointer states in the born-markov approximation},\ }\href@noop {} {\bibfield  {journal} {\bibinfo  {journal} {Phys. Rev. Lett.}\ }\textbf {\bibinfo {volume} {132}},\ \bibinfo {pages} {030203} (\bibinfo {year} {2024})}\BibitemShut {NoStop}%
\bibitem [{\citenamefont {Siva}\ \emph {et~al.}(2023)\citenamefont {Siva} \emph {et~al.}}]{siva2023time}%
  \BibitemOpen
  \bibfield  {author} {\bibinfo {author} {\bibfnamefont {K.}~\bibnamefont {Siva}} \emph {et~al.},\ }\bibfield  {title} {\bibinfo {title} {Time-dependent hamiltonian reconstruction using continuous weak measurements},\ }\href@noop {} {\bibfield  {journal} {\bibinfo  {journal} {PRX Quantum}\ }\textbf {\bibinfo {volume} {4}},\ \bibinfo {pages} {040324} (\bibinfo {year} {2023})}\BibitemShut {NoStop}%
\bibitem [{\citenamefont {Han}\ \emph {et~al.}(2021)\citenamefont {Han}, \citenamefont {Glaz}, \citenamefont {Haile},\ and\ \citenamefont {Lai}}]{han2021tomography}%
  \BibitemOpen
  \bibfield  {author} {\bibinfo {author} {\bibfnamefont {C.-D.}\ \bibnamefont {Han}}, \bibinfo {author} {\bibfnamefont {B.}~\bibnamefont {Glaz}}, \bibinfo {author} {\bibfnamefont {M.}~\bibnamefont {Haile}},\ and\ \bibinfo {author} {\bibfnamefont {Y.-C.}\ \bibnamefont {Lai}},\ }\bibfield  {title} {\bibinfo {title} {Tomography of time-dependent quantum hamiltonians with machine learning},\ }\href@noop {} {\bibfield  {journal} {\bibinfo  {journal} {Phys. Rev. A}\ }\textbf {\bibinfo {volume} {104}},\ \bibinfo {pages} {062404} (\bibinfo {year} {2021})}\BibitemShut {NoStop}%
\bibitem [{\citenamefont {Aguiar}\ \emph {et~al.}(2025)\citenamefont {Aguiar}, \citenamefont {Wold}, \citenamefont {Denisov},\ and\ \citenamefont {Ribeiro}}]{aguiar2025quantum}%
  \BibitemOpen
  \bibfield  {author} {\bibinfo {author} {\bibfnamefont {D.}~\bibnamefont {Aguiar}}, \bibinfo {author} {\bibfnamefont {K.}~\bibnamefont {Wold}}, \bibinfo {author} {\bibfnamefont {S.}~\bibnamefont {Denisov}},\ and\ \bibinfo {author} {\bibfnamefont {P.}~\bibnamefont {Ribeiro}},\ }\bibfield  {title} {\bibinfo {title} {Quantum liouvillian tomography},\ }\href@noop {} {\bibfield  {journal} {\bibinfo  {journal} {arXiv preprint arXiv:2504.10393}\ } (\bibinfo {year} {2025})},\ \Eprint {https://arxiv.org/abs/2504.10393} {arXiv:2504.10393 [quant-ph]} \BibitemShut {NoStop}%
\bibitem [{\citenamefont {Kopciuch}\ \emph {et~al.}(2024)\citenamefont {Kopciuch}, \citenamefont {Smolis}, \citenamefont {Miranowicz},\ and\ \citenamefont {Pustelny}}]{Kopciuch2024}%
  \BibitemOpen
  \bibfield  {author} {\bibinfo {author} {\bibfnamefont {M.}~\bibnamefont {Kopciuch}}, \bibinfo {author} {\bibfnamefont {M.}~\bibnamefont {Smolis}}, \bibinfo {author} {\bibfnamefont {A.}~\bibnamefont {Miranowicz}},\ and\ \bibinfo {author} {\bibfnamefont {S.}~\bibnamefont {Pustelny}},\ }\bibfield  {title} {\bibinfo {title} {Optimized optical tomography of quantum states of a room-temperature alkali-metal vapor},\ }\href@noop {} {\bibfield  {journal} {\bibinfo  {journal} {Phys. Rev. A}\ }\textbf {\bibinfo {volume} {109}},\ \bibinfo {pages} {032402} (\bibinfo {year} {2024})}\BibitemShut {NoStop}%
\bibitem [{\citenamefont {Boulant}\ \emph {et~al.}(2003)\citenamefont {Boulant}, \citenamefont {Havel}, \citenamefont {Pravia},\ and\ \citenamefont {Cory}}]{boulant2003robust}%
  \BibitemOpen
  \bibfield  {author} {\bibinfo {author} {\bibfnamefont {N.}~\bibnamefont {Boulant}}, \bibinfo {author} {\bibfnamefont {T.~F.}\ \bibnamefont {Havel}}, \bibinfo {author} {\bibfnamefont {M.~A.}\ \bibnamefont {Pravia}},\ and\ \bibinfo {author} {\bibfnamefont {D.~G.}\ \bibnamefont {Cory}},\ }\bibfield  {title} {\bibinfo {title} {Robust method for estimating the {L}indblad operators of a dissipative quantum process from measurements of the density operator at multiple time points},\ }\href@noop {} {\bibfield  {journal} {\bibinfo  {journal} {Phys. Rev. A}\ }\textbf {\bibinfo {volume} {67}},\ \bibinfo {pages} {042322} (\bibinfo {year} {2003})}\BibitemShut {NoStop}%
\bibitem [{\citenamefont {Zhang}\ \emph {et~al.}(2012)\citenamefont {Zhang}, \citenamefont {Datta}, \citenamefont {Coldenstrodt-Ronge}, \citenamefont {Jin}, \citenamefont {Eisert}, \citenamefont {Plenio},\ and\ \citenamefont {Walmsley}}]{zhang2012recursive}%
  \BibitemOpen
  \bibfield  {author} {\bibinfo {author} {\bibfnamefont {L.}~\bibnamefont {Zhang}}, \bibinfo {author} {\bibfnamefont {A.}~\bibnamefont {Datta}}, \bibinfo {author} {\bibfnamefont {H.~B.}\ \bibnamefont {Coldenstrodt-Ronge}}, \bibinfo {author} {\bibfnamefont {X.-M.}\ \bibnamefont {Jin}}, \bibinfo {author} {\bibfnamefont {J.}~\bibnamefont {Eisert}}, \bibinfo {author} {\bibfnamefont {M.~B.}\ \bibnamefont {Plenio}},\ and\ \bibinfo {author} {\bibfnamefont {I.~A.}\ \bibnamefont {Walmsley}},\ }\bibfield  {title} {\bibinfo {title} {Recursive quantum detector tomography},\ }\href@noop {} {\bibfield  {journal} {\bibinfo  {journal} {New J. Phys.}\ }\textbf {\bibinfo {volume} {14}},\ \bibinfo {pages} {115005} (\bibinfo {year} {2012})}\BibitemShut {NoStop}%
\bibitem [{\citenamefont {Dezhang~Fard}\ \emph {et~al.}(2025)\citenamefont {Dezhang~Fard}, \citenamefont {Kopciuch}, \citenamefont {Sun}, \citenamefont {W{\l}odarczyk},\ and\ \citenamefont {Pustelny}}]{dezhang2025isolating}%
  \BibitemOpen
  \bibfield  {author} {\bibinfo {author} {\bibfnamefont {A.}~\bibnamefont {Dezhang~Fard}}, \bibinfo {author} {\bibfnamefont {M.}~\bibnamefont {Kopciuch}}, \bibinfo {author} {\bibfnamefont {Y.}~\bibnamefont {Sun}}, \bibinfo {author} {\bibfnamefont {P.}~\bibnamefont {W{\l}odarczyk}},\ and\ \bibinfo {author} {\bibfnamefont {S.}~\bibnamefont {Pustelny}},\ }\bibfield  {title} {\bibinfo {title} {Isolating pure quadratic {Z}eeman splitting},\ }\href@noop {} {\bibfield  {journal} {\bibinfo  {journal} {Phys. Rev. Appl.}\ }\textbf {\bibinfo {volume} {23}},\ \bibinfo {pages} {064034} (\bibinfo {year} {2025})}\BibitemShut {NoStop}%
\bibitem [{\citenamefont {Minganti}\ \emph {et~al.}(2019)\citenamefont {Minganti}, \citenamefont {Miranowicz}, \citenamefont {Chhajlany},\ and\ \citenamefont {Nori}}]{minganti2019quantum}%
  \BibitemOpen
  \bibfield  {author} {\bibinfo {author} {\bibfnamefont {F.}~\bibnamefont {Minganti}}, \bibinfo {author} {\bibfnamefont {A.}~\bibnamefont {Miranowicz}}, \bibinfo {author} {\bibfnamefont {R.~W.}\ \bibnamefont {Chhajlany}},\ and\ \bibinfo {author} {\bibfnamefont {F.}~\bibnamefont {Nori}},\ }\bibfield  {title} {\bibinfo {title} {Quantum exceptional points of non-{H}ermitian {H}amiltonians and {L}iouvillians: The effects of quantum jumps},\ }\href@noop {} {\bibfield  {journal} {\bibinfo  {journal} {Phys. Rev. A}\ }\textbf {\bibinfo {volume} {100}},\ \bibinfo {pages} {062131} (\bibinfo {year} {2019})}\BibitemShut {NoStop}%
\bibitem [{\citenamefont {Kopciuch}\ and\ \citenamefont {Miranowicz}(2025)}]{kopciuch2025liouvillian}%
  \BibitemOpen
  \bibfield  {author} {\bibinfo {author} {\bibfnamefont {M.}~\bibnamefont {Kopciuch}}\ and\ \bibinfo {author} {\bibfnamefont {A.}~\bibnamefont {Miranowicz}},\ }\bibfield  {title} {\bibinfo {title} {Liouvillian and hamiltonian exceptional points of atomic vapors: The spectral signatures of quantum jumps},\ }\href@noop {} {\bibfield  {journal} {\bibinfo  {journal} {Phys. Rev. Res.}\ }\textbf {\bibinfo {volume} {7}},\ \bibinfo {pages} {033187} (\bibinfo {year} {2025})}\BibitemShut {NoStop}%
\bibitem [{\citenamefont {Abo}\ \emph {et~al.}(2024)\citenamefont {Abo}, \citenamefont {Tulewicz}, \citenamefont {Bartkiewicz}, \citenamefont {{\"O}zdemir},\ and\ \citenamefont {Miranowicz}}]{abo2024experimental}%
  \BibitemOpen
  \bibfield  {author} {\bibinfo {author} {\bibfnamefont {S.}~\bibnamefont {Abo}}, \bibinfo {author} {\bibfnamefont {P.}~\bibnamefont {Tulewicz}}, \bibinfo {author} {\bibfnamefont {K.}~\bibnamefont {Bartkiewicz}}, \bibinfo {author} {\bibfnamefont {{\c{S}}.~K.}\ \bibnamefont {{\"O}zdemir}},\ and\ \bibinfo {author} {\bibfnamefont {A.}~\bibnamefont {Miranowicz}},\ }\bibfield  {title} {\bibinfo {title} {Experimental {L}iouvillian exceptional points in a quantum system without {H}amiltonian singularities},\ }\href@noop {} {\bibfield  {journal} {\bibinfo  {journal} {New J. Phys.}\ }\textbf {\bibinfo {volume} {26}},\ \bibinfo {pages} {123032} (\bibinfo {year} {2024})}\BibitemShut {NoStop}%
\bibitem [{\citenamefont {Wolf}\ \emph {et~al.}(2008)\citenamefont {Wolf}, \citenamefont {Eisert}, \citenamefont {Cubitt},\ and\ \citenamefont {Cirac}}]{wolf2008assessing}%
  \BibitemOpen
  \bibfield  {author} {\bibinfo {author} {\bibfnamefont {M.~M.}\ \bibnamefont {Wolf}}, \bibinfo {author} {\bibfnamefont {J.}~\bibnamefont {Eisert}}, \bibinfo {author} {\bibfnamefont {T.~S.}\ \bibnamefont {Cubitt}},\ and\ \bibinfo {author} {\bibfnamefont {J.~I.}\ \bibnamefont {Cirac}},\ }\bibfield  {title} {\bibinfo {title} {Assessing non-markovian quantum dynamics},\ }\href@noop {} {\bibfield  {journal} {\bibinfo  {journal} {Phys. Rev. Lett.}\ }\textbf {\bibinfo {volume} {101}},\ \bibinfo {pages} {150402} (\bibinfo {year} {2008})}\BibitemShut {NoStop}%
\bibitem [{\citenamefont {Rivas}\ \emph {et~al.}(2014)\citenamefont {Rivas}, \citenamefont {Huelga},\ and\ \citenamefont {Plenio}}]{Rivas2014Quantum}%
  \BibitemOpen
  \bibfield  {author} {\bibinfo {author} {\bibfnamefont {A.}~\bibnamefont {Rivas}}, \bibinfo {author} {\bibfnamefont {S.~A.}\ \bibnamefont {Huelga}},\ and\ \bibinfo {author} {\bibfnamefont {M.~B.}\ \bibnamefont {Plenio}},\ }\bibfield  {title} {\bibinfo {title} {Quantum non-markovianity: characterization, quantification and detection},\ }\href {https://doi.org/10.1088/0034-4885/77/9/094001} {\bibfield  {journal} {\bibinfo  {journal} {Rep. Prog. Phys.}\ }\textbf {\bibinfo {volume} {77}},\ \bibinfo {pages} {094001} (\bibinfo {year} {2014})}\BibitemShut {NoStop}%
\bibitem [{\citenamefont {Luan}\ \emph {et~al.}(2024)\citenamefont {Luan}, \citenamefont {Li}, \citenamefont {Zheng}, \citenamefont {Kuang}, \citenamefont {Yu},\ and\ \citenamefont {Zhang}}]{Luan2024Quantum}%
  \BibitemOpen
  \bibfield  {author} {\bibinfo {author} {\bibfnamefont {T.}~\bibnamefont {Luan}}, \bibinfo {author} {\bibfnamefont {Z.}~\bibnamefont {Li}}, \bibinfo {author} {\bibfnamefont {C.}~\bibnamefont {Zheng}}, \bibinfo {author} {\bibfnamefont {X.}~\bibnamefont {Kuang}}, \bibinfo {author} {\bibfnamefont {X.}~\bibnamefont {Yu}},\ and\ \bibinfo {author} {\bibfnamefont {Z.}~\bibnamefont {Zhang}},\ }\bibfield  {title} {\bibinfo {title} {Non-{M}arkovian {Q}uantum {P}rocess {T}omography},\ }\href {https://doi.org/https://doi.org/10.3390/sym16020180} {\bibfield  {journal} {\bibinfo  {journal} {Symmetry}\ }\textbf {\bibinfo {volume} {16}},\ \bibinfo {pages} {180} (\bibinfo {year} {2024})}\BibitemShut {NoStop}%
\bibitem [{\citenamefont {Browaeys}\ and\ \citenamefont {Lahaye}(2020)}]{Browaeys2020Many}%
  \BibitemOpen
  \bibfield  {author} {\bibinfo {author} {\bibfnamefont {A.}~\bibnamefont {Browaeys}}\ and\ \bibinfo {author} {\bibfnamefont {T.}~\bibnamefont {Lahaye}},\ }\bibfield  {title} {\bibinfo {title} {Many-body physics with individually controlled {R}ydberg atoms},\ }\href@noop {} {\bibfield  {journal} {\bibinfo  {journal} {Nat. Phys.}\ }\textbf {\bibinfo {volume} {16}},\ \bibinfo {pages} {132} (\bibinfo {year} {2020})}\BibitemShut {NoStop}%
\bibitem [{\citenamefont {Joas}\ \emph {et~al.}(2025)\citenamefont {Joas} \emph {et~al.}}]{joas2025high}%
  \BibitemOpen
  \bibfield  {author} {\bibinfo {author} {\bibfnamefont {T.}~\bibnamefont {Joas}} \emph {et~al.},\ }\bibfield  {title} {\bibinfo {title} {High-fidelity electron spin gates for scaling diamond quantum registers},\ }\href@noop {} {\bibfield  {journal} {\bibinfo  {journal} {Phys. Rev. X}\ }\textbf {\bibinfo {volume} {15}},\ \bibinfo {pages} {021069} (\bibinfo {year} {2025})}\BibitemShut {NoStop}%
\bibitem [{\citenamefont {Dai}\ \emph {et~al.}(2021)\citenamefont {Dai} \emph {et~al.}}]{dai2021calibration}%
  \BibitemOpen
  \bibfield  {author} {\bibinfo {author} {\bibfnamefont {X.}~\bibnamefont {Dai}} \emph {et~al.},\ }\bibfield  {title} {\bibinfo {title} {Calibration of flux crosstalk in large-scale flux-tunable superconducting quantum circuits},\ }\href@noop {} {\bibfield  {journal} {\bibinfo  {journal} {PRX Quantum}\ }\textbf {\bibinfo {volume} {2}},\ \bibinfo {pages} {040313} (\bibinfo {year} {2021})}\BibitemShut {NoStop}%
\bibitem [{\citenamefont {Blatt}\ and\ \citenamefont {Roos}(2012)}]{Blatt2012Quantum}%
  \BibitemOpen
  \bibfield  {author} {\bibinfo {author} {\bibfnamefont {R.}~\bibnamefont {Blatt}}\ and\ \bibinfo {author} {\bibfnamefont {C.}~\bibnamefont {Roos}},\ }\bibfield  {title} {\bibinfo {title} {Quantum simulations with trapped ions},\ }\href@noop {} {\bibfield  {journal} {\bibinfo  {journal} {Nat. Phys.}\ }\textbf {\bibinfo {volume} {8}},\ \bibinfo {pages} {277} (\bibinfo {year} {2012})}\BibitemShut {NoStop}%
\bibitem [{\citenamefont {Torlai}\ \emph {et~al.}(2018)\citenamefont {Torlai} \emph {et~al.}}]{Torlai2018Neutral}%
  \BibitemOpen
  \bibfield  {author} {\bibinfo {author} {\bibfnamefont {H.}~\bibnamefont {Torlai}} \emph {et~al.},\ }\bibfield  {title} {\bibinfo {title} {Neural-network quantum state tomography},\ }\href@noop {} {\bibfield  {journal} {\bibinfo  {journal} {Nat. Phys.}\ }\textbf {\bibinfo {volume} {14}},\ \bibinfo {pages} {447} (\bibinfo {year} {2018})}\BibitemShut {NoStop}%
\bibitem [{\citenamefont {Zhang}(2025)}]{Zhang2025Learning}%
  \BibitemOpen
  \bibfield  {author} {\bibinfo {author} {\bibfnamefont {X.~t.}\ \bibnamefont {Zhang}},\ }\bibfield  {title} {\bibinfo {title} {Learning and forecasting open quantum dynamics with correlated noise},\ }\href@noop {} {\bibfield  {journal} {\bibinfo  {journal} {Commun. Phys.}\ }\textbf {\bibinfo {volume} {8}},\ \bibinfo {pages} {29} (\bibinfo {year} {2025})}\BibitemShut {NoStop}%
\bibitem [{\citenamefont {Bukov}\ \emph {et~al.}(2018)\citenamefont {Bukov} \emph {et~al.}}]{Bukov2018Reinforcement}%
  \BibitemOpen
  \bibfield  {author} {\bibinfo {author} {\bibfnamefont {M.}~\bibnamefont {Bukov}} \emph {et~al.},\ }\bibfield  {title} {\bibinfo {title} {Reinforcement learning in different phases of quantum control},\ }\href@noop {} {\bibfield  {journal} {\bibinfo  {journal} {Phys. Rev. X}\ }\textbf {\bibinfo {volume} {8}},\ \bibinfo {pages} {031086} (\bibinfo {year} {2018})}\BibitemShut {NoStop}%
\bibitem [{\citenamefont {Niu}\ \emph {et~al.}(2019)\citenamefont {Niu} \emph {et~al.}}]{Niu2021Universal}%
  \BibitemOpen
  \bibfield  {author} {\bibinfo {author} {\bibfnamefont {M.~Y.}\ \bibnamefont {Niu}} \emph {et~al.},\ }\bibfield  {title} {\bibinfo {title} {Universal quantum control through deep reinforcement learning},\ }\href@noop {} {\bibfield  {journal} {\bibinfo  {journal} {npj Quantum Inf.}\ }\textbf {\bibinfo {volume} {5}},\ \bibinfo {pages} {33} (\bibinfo {year} {2019})}\BibitemShut {NoStop}%
\end{thebibliography}%

\end{document}


\newcommand{\JUAddress}{Marian Smoluchowski Institute of Physics, Jagiellonian University in Krak\'ow, 30-348 Krak\'ow, Poland}
\newcommand{\JUDSAddress}{Doctoral School of Exact and Natural Sciences, Jagiellonian University in Krak\'ow, 30-348 Krak\'ow, Poland}
\newcommand{\UAMAddress}{Institute of Spintronics and Quantum Information, Faculty of Physics, Adam Mickiewicz University, 61-614 Pozna\'n, Poland}
\newcommand{\HUaddress}{Department of Physics, Harvard University, Cambridge, MA 02138, USA}
\newcommand{\ZJUaddress}{School of Physics, Zhejiang University, Hangzhou 310027, China}

\author{Yujie Sun}
\affiliation{\JUAddress}
\affiliation{\ZJUaddress}

\author{Marek Kopciuch}
\email{marek.kopciuch@amu.edu.pl}
\affiliation{\UAMAddress}
\affiliation{\JUAddress}

\author{Arash Dezhang Fard}
\affiliation{\JUAddress}
\affiliation{\JUDSAddress}

\author{Szymon Pustelny}
\email{szymon.pustelny@uj.edu.pl}
\affiliation{\JUAddress}
\affiliation{\HUaddress}


\title{Supplementary Information for Quantum Process Tomography of a Thermal Alkali-Metal Vapor}


\author{}
\affiliation{}


\date{\today}



\maketitle
\section{Quantum-state vectorization in the Bloch--Fano representation}

The density matrix $\hat{\rho}$, describing a quantum system with $d$ energy levels, is expanded in the Bloch--Fano representation \cite{bertlmann2008bloch, Benenti2009}:
\begin{equation}
    \hat{\rho} = \frac{1}{d}\hat{\mathds{1}}_d+\sum_{i=1}^{d^{2}-1}\rho_{i}\hat{\sigma}_{i}=\sum_{i=1}^{d^2}\rho_{i}\hat{\sigma}_{i},
    \label{eq:project}
\end{equation}
where $\hat{\sigma}_i$ for $i\in\{1,\dots,d^2-1\}$ are the generalized Pauli matrices corresponding to the generators of SU($d$). In addition, the operator $\hat{\sigma}_{d^2}=\sqrt{2/d}\hat{\mathds{1}}_d$, where $\hat{\mathds{1}}_d$ is the $d$-dimensional identity matrix, is introduced to express the expansion in a compact form. The scaling factor is chosen such that the basis operators satisfy the orthogonality relation:
\begin{equation}
    \frac{1}{2}\Tr[\hat{\sigma}_{i}\hat{\sigma}_{j}]=\delta_{ij}.
\end{equation}
The corresponding coefficients $\rho_i$, representing the projections of the density matrix onto the basis operators $\hat{\sigma}_i$, are given by:
\begin{equation}
    \rho_{i}=\frac{1}{2}\Tr(\hat{\rho}\hat{\sigma}_{i}).
\end{equation}
This mapping defines an isomorphism between the density operator and its vectorized representation in Liouville space:
\begin{equation}
    \hat{\rho} = \sum_{i=1}^{d^2} \frac{1}{2}\Tr(\hat{\rho}\hat{\sigma}_{i}) \hat{\sigma}_{i} = \sket{\rho} \Longleftrightarrow \sket{\rho}_{i} = \rho_{i}= \frac{1}{2}\Tr(\hat{\rho} \hat{\sigma}_i),
\end{equation}
where the coefficients are real-valued. The final coefficient is constrained by the trace conservation condition $\rho_{d^2}=\sqrt{1/(2d)}$. Consequently, a general density matrix $\hat{\rho}$ is uniquely mapped as a real $d^2$-dimensional state vector $\sket{\rho}$ in the Bloch--Fano representation.

To complete the framework, we map the Liouvillian onto its matrix representation $\dhat{L}$ acting in operator space:
\begin{equation}
    \mathcal{L}[\hat{\rho}] \rightarrow \dhat{L} \sket{\rho} = -\left(i \dhat{H} + \dhat{R} \right)\sket{\rho},
    \label{eq:sup_liouvillian}
\end{equation}
where the coherent part is expanded as:
\begin{equation}
\begin{split}
    \comm{\hat{H}}{\hat{\rho}} &= \sum_{i=1}^{d^2} \comm{\hat{H}}{\hat{\sigma}_i}\rho_i = \sum_{i,j=1}^{d^2}  \dfrac{1}{2} \Tr(\comm{\hat{H}}{\hat{\sigma}_{i}} \hat{\sigma}_j)  \rho_i \hat{\sigma}_j\\
    &= \sum_{i,j=1}^{d^2}\hat{\sigma}_{j}\dhat{H}_{ji} \sket{\rho}_i = \dhat{H}\sket{\rho},
\end{split}
\end{equation}
with $\dhat{H}$ being a matrix with elements:
\begin{equation}
    \dhat{H}_{ij} = \dfrac{1}{2} \Tr(\comm{\hat{H}}{\hat{\sigma}_j}\hat{\sigma}_i).
    \label{eq:sup_H}
\end{equation}
Similarly, the dissipative part in Eq.~\eqref{eq:sup_liouvillian} is expanded as:
\begin{eqnarray}
    &&\sum_{\mu} \left( \hat{L}_{\mu} \hat{\rho} \hat{L}_{\mu}^{\dagger} - \dfrac{1}{2}\acomm{\hat{L}_{\mu}^{\dagger}\hat{L}_{\mu}}{\hat{\rho}} \right) = \nonumber \\
    &&=\sum_{i,j=1}^{d^2} \dfrac{1}{2}\sum_{\mu} \Tr( \left[ \hat{L}_{\mu} \hat{\sigma}_i \hat{L}_{\mu}^{\dagger} - \frac{1}{2}\acomm{\hat{L}_{\mu}^{\dagger}\hat{L}_{\mu}}{\hat{\sigma}_i} \right]\hat{\sigma}_j) \rho_{i}\hat{\sigma}_j \\
   &&=\sum_{i,j=1}^{d^2} \hat{\sigma}_{j} (-\dhat{R}_{ji}) \sket{\rho}_{i} = -\dhat{R} \sket{\rho}, \nonumber
\end{eqnarray}
with:
\begin{equation}
    \dhat{R}_{ij} = \dfrac{1}{2}\sum_{\mu} \Tr( \left[\frac{1}{2}\acomm{\hat{L}_{\mu}^{\dagger}\hat{L}_{\mu}}{\hat{\sigma}_j} - \hat{L}_{\mu} \hat{\sigma}_j \hat{L}_{\mu}^{\dagger} \right]\hat{\sigma}_i).
    \label{eq:R_matrix}
\end{equation}
Equations~\eqref{eq:sup_liouvillian}--\eqref{eq:R_matrix} fully express the Lindblad master equation in the Bloch--Fano representation, underpinning the tomography protocol.

For our experimental implementation, we focus on a qutrit, realized in the $f=1$ hyperfine ground-state manifold of the $^{87}$Rb atom. In this case, the generalized Pauli matrices are given by the Gell–Mann matrices, supplemented by $\hat{\sigma}_{9} = \sqrt{2/3}\hat{\mathds{1}}_3$:
\begin{equation}
    \begin{array}{ccc}
        \hat{\sigma}_{1} = \left (
        \begin{matrix}
            0 & 1 & 0\\
            1 & 0 & 0\\
            0 & 0 & 0\\
        \end{matrix}\right ) &
        \hat{\sigma}_{2} = \left (
        \begin{matrix}
            0 & -i & 0\\
            i & 0 & 0\\
            0 & 0 & 0\\
        \end{matrix}\right ) &
        \hat{\sigma}_{3} = \left (
        \begin{matrix}
            1 & 0 & 0\\
            0 & -1 & 0\\
            0 & 0 & 0\\
        \end{matrix}\right ) \\
        \\
        \hat{\sigma}_{4} = \left (
        \begin{matrix}
            0 & 0 & 1\\
            0 & 0 & 0\\
            1 & 0 & 0\\
        \end{matrix}\right ) &
        \hat{\sigma}_{5} = \left (
        \begin{matrix}
            0 & 0 & -i\\
            0 & 0 & 0\\
            i & 0 & 0\\
        \end{matrix}\right ) &
        \hat{\sigma}_{6} = \left (
        \begin{matrix}
            0 & 0 & 0\\
            0 & 0 & 1\\
            0 & 1 & 0\\
        \end{matrix}\right ) \\
        \\
        \hat{\sigma}_{7} = \left (
        \begin{matrix}
            0 & 0 & 0\\
            0 & 0 & -i\\
            0 & i & 0\\
        \end{matrix}\right ) &
        \hat{\sigma}_{8} = \frac{1}{\sqrt{3}}\left (
        \begin{matrix}
            1 & 0 & 0\\
            0 & 1 & 0\\
            0 & 0 & -2\\
        \end{matrix}\right ) &
        \hat{\sigma}_{9} = \sqrt{\frac{2}{3}}\left (
        \begin{matrix}
            1 & 0 & 0\\
            0 & 1 & 0\\
            0 & 0 & 1\\
        \end{matrix}\right ) \\
    \end{array}
\end{equation}
Thus, $\sket{\rho}$ is a 9-dimensional state vector, and the process matrix $\dhat{P}(t)$, which accounts for the evolution of the atomic system, is a $9\times 9$ real matrix.

\section{Effective dissipation\label{sec:effective_dissipation}}

Residual, slowly varying magnetic fields are unavoidable in most atomic-vapor experiments, arising from imperfect shielding or environmental fluctuations. Although these fields enter the dynamics through a Hamiltonian term, it is convenient to absorb their effect into an effective relaxation superoperator. In Liouville space, this represents a redistribution of the generator components, but translating the result back to operator language requires a new set of jump operators that reproduce the same evolution. The construction is performed within the Gorini–Kossakowski–Lindblad–Sudarshan (GKLS) formalism \cite{gorini1976completely}.

Working in the operator basis $\{\hat{\sigma}_{1},\dots,\hat{\sigma}_{d^2}\}$, the master equation may be written as:
\begin{equation}
    \dv{\hat{\rho}}{t} = -i \comm{\hat{H}_C(t)}{\hat{\rho}} -i \comm{\hat{H}_R}{\hat{\rho}} - \frac{1}{2}\sum_{i,j=1}^{d^2}\mathbb{C}_{ij}\left( \comm{\hat{\sigma}_{i}}{\hat{\rho}\hat{\sigma}_{j}^{\dagger}} + \comm{\hat{\sigma}_{i}\hat{\rho}}{\hat{\sigma}_{j}^{\dagger}}\right),
    \label{eq:GKLS}
\end{equation}
where $\mathbb{C}_{ij}$ are the (generally complex) elements of the Kossakowski matrix $\mathbb{C}$. Unlike the standard formulation, the sum includes the basis element proportional to the identity, $\hat{\sigma}_{d^{2}}=\sqrt{2/d}\,\hat{\mathds{1}}_d$, which is required for the redistribution of coherent terms.

By expanding the residual magnetic-field Hamiltonian $\hat{H}_{R}$ in the generalized Pauli matrices, $\hat{H}_{R}=\sum_{k}h_{k}\hat{\sigma}_{k}$, where $h_{k}=\tfrac12\mathrm{Tr}(\hat{H}_{R}\hat{\sigma}_{k})$, we express the second term in Eq.~\eqref{eq:GKLS} as:
\begin{equation}
    \begin{split}
        -i\comm{\hat{H}_R}{\hat{\rho}}&=-i\sum_{k}h_{k}\,[\hat{\sigma}_{k},\hat{\rho}]\\
        &=-\frac{i\sqrt{d}}{\sqrt{2}}\sum_{k}h_{k}\bigl([\hat{\sigma}_{k},\hat{\rho}\hat{\sigma}_{d^{2}}]+[\hat{\sigma}_{k}\hat{\rho},\hat{\sigma}_{d^{2}}]\bigr).
    \end{split}
    \label{eq:Hamiltonian_as_dissipator}
\end{equation}
Comparison between Eqs.~\eqref{eq:GKLS} and \eqref{eq:Hamiltonian_as_dissipator} shows that the Hamiltonian contribution can be absorbed into the dissipator through the replacement:
\begin{equation}
    \label{eq:C_shift}
    \mathbb{C}_{ij}\;\longrightarrow\; \mathbb{C}'_{ij}= \mathbb{C}_{ij}+\frac{i\sqrt{d}}{\sqrt{2}}\;\bigl(h_{i}\delta_{j,d^{2}}-h_j\delta_{i,d^{2}}\bigr).
\end{equation}

The transformation in Eq.~\eqref{eq:C_shift} couples the coherent and dissipative parts of the generator and may produce a modified Kossakowski matrix $\mathbb{C}'$ that is not positive semi-definite.[1] This preserves the completely positive and trace-preserving (CPTP) nature of the overall Liouvillian, as we have only redistributed terms without altering the Liouvillian's action on $\hat{\rho}$.

Equation \eqref{eq:R_matrix} expresses the dissipative contribution in terms of the microscopic jump operators $\hat{L}_{\mu}$ and is therefore general. For the Bloch--Fano representation used in the present reconstruction, however, it is more convenient to rewrite the dissipator directly in the operator basis $\hat{\sigma}_{i}$. To clarify how the elements of the modified Kossakowski matrix $\mathbb{C}'$ determine the effective relaxation superoperator $\dhat{R}_T$, we introduce the symmetric ($d_{ijk}$) and antisymmetric ($f_{ijk}$) structure constants:
\begin{equation}
    \label{eq: symmetry}
    d_{ijk}=\frac{1}{4}\Tr\left(\hat{\sigma}_{i}\acomm{\hat{\sigma}_{j}}{\hat{\sigma}_{k}}\right),\quad f_{ijk}=-\frac{i}{4}\Tr\left(\hat{\sigma}_{i}[\hat{\sigma}_{j},\hat{\sigma}_{k} ]\right).
\end{equation}
The operator basis therefore satisfies:
\begin{equation}
    \label{eq: antisymmetry}
    [\hat{\sigma}_{i},\hat{\sigma}_{j}] =2i\sum_{k=1}^{d^2}f_{ijk}\hat{\sigma}_{k},\quad\acomm{\hat{\sigma}_{i}}{\hat{\sigma}_{j}} =2\sum_{k=1}^{d^2}d_{ijk}\hat{\sigma}_{k},\quad \hat{\sigma}_{i}\hat{\sigma}_{j}=\sum_{k=1}^{d^2}(if_{ijk}+d_{ijk})\hat{\sigma}_{k}. 
\end{equation}
Expanding the jump operators in the Bloch--Fano basis and substituting the result into Eq. \eqref{eq:R_matrix}, we obtain the matrix elements of the effective relaxation superoperator:
\begin{equation}
    \begin{aligned}
        (\dhat{R}_T)_{ij}
        &= \frac{1}{2}\sum_{m,n} \mathbb{C}'_{mn}
        \Tr\!\Big(
        \big[\tfrac{1}{2}\acomm{\hat{\sigma}_n\hat{\sigma}_m}{\hat{\sigma}_j}
        - \hat{\sigma}_m \hat{\sigma}_j \hat{\sigma}_n\big]\hat{\sigma}_i
        \Big) \\
        &= -\frac{i}{2}\sum_{m,n} \mathbb{C}'_{mn}
        \Tr\!\Bigg(\left[\sum_{k} (f_{mjk}\hat{\sigma}_{k}\hat{\sigma}_{n}+f_{mnk}\hat{\sigma}_{j}\hat{\sigma}_{k}+f_{jnk}\hat{\sigma}_{m}\hat{\sigma}_{k}+f_{mnk}\hat{\sigma}_{k}\hat{\sigma}_{j})\right]\hat{\sigma}_{i}
        \Bigg) \\
        &= -i\sum_{m,n,k} \mathbb{C}'_{mn}\left[ f_{mjk}(d_{kni}+if_{kni})+2f_{mnk}d_{jki}+f_{jnk}(d_{mki}+if_{mki})\right].
    \end{aligned}    
\end{equation}
This relation provides an explicit mapping between the modified Kossakowski matrix $\mathbb{C}'$ and the effective relaxation superoperator $\dhat{R}_T$. In the numerical reconstruction, the elements of $\mathbb{C}'$ are treated as free parameters and optimized such that the resulting superoperator $\dhat{R}_T$ yields the best agreement with the experimentally reconstructed process matrix. 

In practice, we first identify and remove the coherent Hamiltonian contribution associated with residual external fields from the reconstructed generator. The remaining non-unitary component is then assigned to the dissipative sector and represented by the Kossakowski matrix $\mathbb{C}$. Since this decomposition is performed on a numerically reconstructed generator, the resulting matrix $\mathbb{C}$ is not guaranteed to be positive semi-definite. In particular, the unconstrained reconstruction may produce small negative eigenvalues $\lambda_k<0$, which are incompatible with a physically admissible GKLS dissipator. After diagonalization of $\mathbb{C}$, the corresponding jump operators are defined as:
\begin{equation}
\hat{L}_{k}=\sqrt{\lambda_k}\sum_j U_{jk}\hat{\sigma}_j,
\end{equation}
where $U$ denotes the unitary matrix defining the eigenbasis of $\mathbb{C}$. Negative $\lambda_k$ therefore corresponds to unphysical decoherence rates. To enforce physical admissibility, we applied spectral regularization to the reconstructed Kossakowski matrix post hoc. Specifically, negative eigenvalues are truncated to zero, and the remaining positive eigenvalues are rescaled to preserve the total trace of the dissipative sector.

The final dissipative contribution reported in this work is obtained from the regularized positive semi-definite Kossakowski matrix.

\section{Optical Configuration}

As described in the main text, state preparation and measurement are driven by a set of three independently stabilized lasers acting on the D$_1$ line of $^{87}$Rb.

The pump beam, tuned to the atomic $f=1 \rightarrow F=1$ transition, is generated by a Fabry-Pérot extended-cavity diode laser (ECDL) and is used to initialize the atomic population via resonant optical pumping into the target manifold. To generate a specific quantum state, the pump light propagates along the $x$-axis and is prepared in a circular polarization state through a Glan-Thompson polarizer and a quarter-wave plate.

At the same time, a Ti:Sapphire repump laser, tuned to the atomic transition $f=2 \rightarrow F=2$, operates continuously to prevent population trapping in dark hyperfine states. This provides the physical basis for the strict trace-preservation constraint employed in our reconstruction. The repump light also propagates along the $x$-axis but is linearly polarized along the $z$-axis during the interaction with the atoms.

Finally, an off-resonant probe beam, generated by a distributed-feedback (DFB) diode laser, passes through the ensemble along the $z$-axis and is linearly polarized along the $y$-axis. To minimize probe-induced back-action, its intensity is passively attenuated with high-quality crystal polarizers and fixed at \SI{10}{\micro\watt/\square{\centi\meter}}. Its frequency is blue-detuned by \SI{50}{\mega\hertz} from the atomic transition $f=1\rightarrow F=2$. The resulting Faraday rotation, mapped from the macroscopic atomic coherence, is measured with a balanced polarimeter (consisting of a Wollaston prism and a balanced photodetector) to record the time-resolved free precession decay (FPD).

The frequencies of all three lasers are independently controlled and stabilized. The pump and probe lasers are locked using a dichroic atomic vapor laser lock (DAVLL), whereas the repump laser is stabilized by an internal automatic locking module. Laser frequencies are monitored via a wavemeter, with pump and probe beams further referenced to saturated-absorption spectroscopy. The beam intensities and pulse timings are controlled by three independent acousto-optic modulators (AOMs), ensuring that all laser fields are fully switched off during the application of the state-preparation magnetic-field pulses.

\section{Quantum state tomography of a qutrit system\label{sec:QST}}

Quantum State Tomography (QST) enables the reconstruction of the density matrix through a series of informationally complete measurements. The QST technique used in this work is described in detail in Ref.~\cite{Kopciuch2024}.

When a linearly polarized optical beam propagates through an atomic vapor, its polarization state undergoes measurable perturbations determined by the atomic susceptibility tensor. If the atomic system exhibits time-dependent quantum dynamics, these parameters are encoded into the probe beam's Stokes parameters depending on time. In the absence of external driving, the dynamics are governed by intrinsic relaxation. This constitutes time-dependent FPD signals.  

By detecting the FPD signals originating from the atomic ensemble, we can access the time-resolved evolution of atomic coherence and population imbalance. As shown in Ref.~\cite{Kopciuch2024}, the time- and detuning-dependent polarization rotation $\delta\alpha$ is given by:
\begin{equation}
    \delta\alpha(t;\Delta\nu)= \eta(\Delta\nu)\left(e ^ { - \gamma_2 t } \left[\left\langle\hat{\alpha}_R\right\rangle \sin \left(2 \Omega_L t\right)+\left\langle\hat{\alpha}_I\right\rangle \cos \left(2 \Omega_L t\right)\right]-\zeta(\Delta\nu) e^{-\gamma_1 t}\langle\hat{\beta}\rangle\right)
\end{equation}
where $\eta(\Delta\nu)$ and $\zeta(\Delta\nu)$ are the global and local scaling factors, respectively; the operators $\hat{\alpha}_{R,I}$ and $\hat{\beta}$ represent the coherence and population difference between the levels differing in magnetic quantum number $m$ by 2. For our system, probed on the $f=1\rightarrow F=2$ transition, these operators are defined as:
\begin{equation}
    \begin{aligned}
        \hat{\alpha}_{R} &= \frac{1}{30}\left(\ket{-1}\bra{1}+\ket{1}\bra{-1}\right),\\
        \hat{\alpha}_{I} &= \frac{i}{30}\left(\ket{-1}\bra{1}-\ket{1}\bra{-1}\right),\\
        \hat{\beta}&=\frac{1}{6}\left(\ket{-1}\bra{-1}-\ket{1}\bra{1}\right).
    \end{aligned}
\end{equation}
The polarization rotation $\delta\alpha$ is also determined by the Larmor frequency $\Omega_{L}$ and the longitudinal and transverse relaxation rates $\gamma_{1}$ and $\gamma_{2}$. To suppress technical noise and systematic errors, we adopt the cyclically ordered phase sequence (CYCLOPS) method \cite{Kopciuch2024}. 

\begin{figure}[htbp]
    \centering
    \includegraphics[width=1\linewidth]{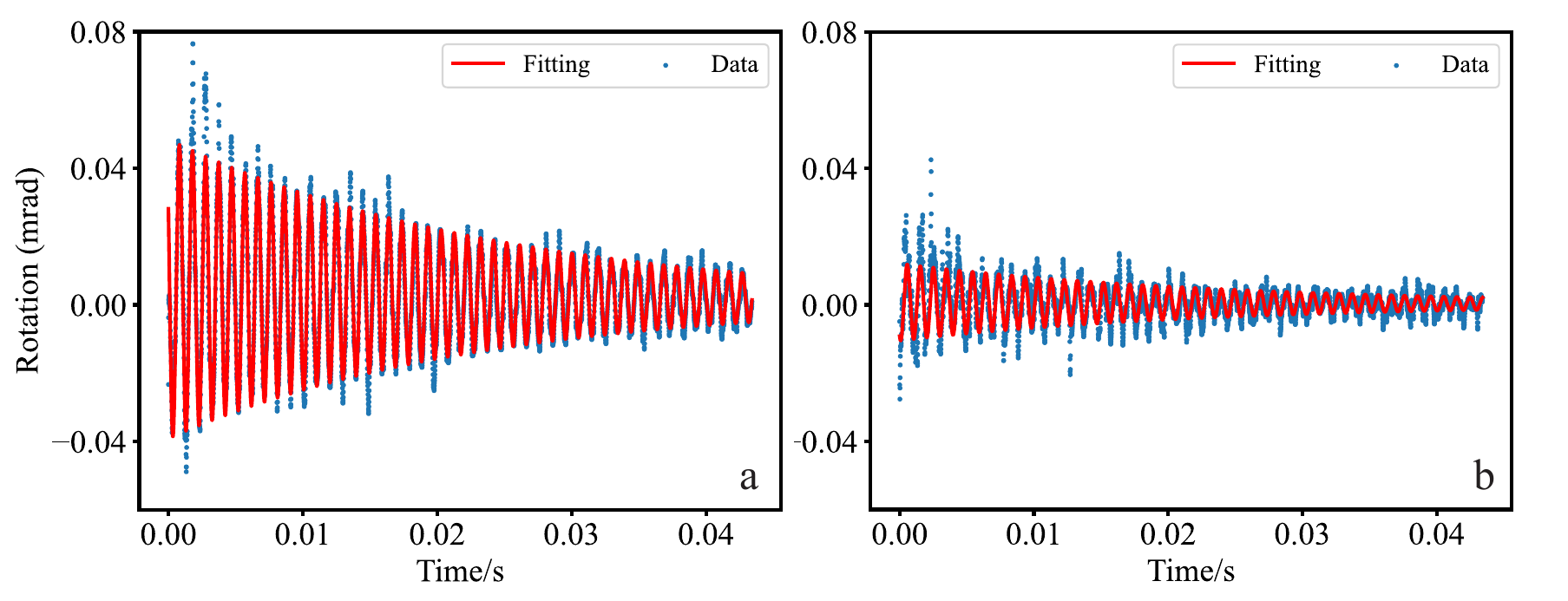}
    \caption{Detected polarization rotations (blue dots) and fitted traces (red lines) obtained from CYCLOPS for a qutrit state with maximal stretch along the $x$-axis. (a) The fitting function is $\delta \alpha = \eta \big[ \langle \hat{\alpha}_R \rangle \sin(2\Omega_L t + \varphi) e^{-\gamma_1 t} - \langle \hat{\beta} \rangle \zeta e^{-\gamma_2 t} \big]$, yielding $\langle\hat{\beta}\rangle=4.26\times10^{-3}$ and $\langle\hat{\alpha}_{R}\rangle=1.56\times10^{-2}$. (b) The fitting function is $\delta \alpha = \eta \big[ \langle \hat{\alpha}_I \rangle \cos(2\Omega_L t + \varphi) e^{-\gamma_1 t} - \langle \hat{\beta} \rangle \zeta e^{-\gamma_2 t} \big]$, yielding $\langle\hat{\beta}\rangle=7.12\times10^{-4}$ and $\langle\hat{\alpha}_{I}\rangle=-4.03\times10^{-3}$. }
    \label{fig:FiD}
\end{figure}

Figures~\ref{fig:FiD}(a) and (b) show the experimentally measured polarization rotations (blue dots) and the corresponding theoretical fits (red lines), obtained via a CYCLOPS measurement for a qutrit polarized along the $x$-axis. This measurement captures the state-dependent polarization response in a single measurement basis, but additional bases are required to access the full set of independent qutrit degrees of freedom.

We therefore perform a set of complementary measurements by applying $\pi/2$ control pulses around the $x$- and $y$-axes \cite{Kopciuch2024}. These rotations project different density-matrix components onto the measured observable and thus provide the linearly independent constraints required for tomographic completeness. Finally, we employ a linear inversion method to obtain an initial estimate, followed by a maximum likelihood estimation to find the closest physical realization of the full density matrix $\hat{\rho}$.

\section{Quantum process tomography --- dissipator reconstruction}

As established in the main text, the Quantum Process Tomography (QPT) framework relies on comparing a set of initial states with their counterparts after undergoing the measured process. For complete QPT characterization, we prepare an informationally complete set of initial states spanning the Hilbert space of a qutrit system. This set consists of $N=15$ linearly independent states, constructed as eigenstates of the Gell-Mann matrices:
\begin{equation}
    \ket{\psi_{k}} = \left\{ 
    \begin{array}{l}
         \ket{i} \\
         \frac{1}{\sqrt{2}}\left(\ket{i} + e^{i\phi}\ket{j}\right)
    \end{array}
    \right.
    \label{ini_states}
\end{equation}
where $i, j \in \{-1, 0, 1\}$ index the qutrit levels, and $\phi \in \{0, \pi/2, \pi, 3\pi/2\}$ ensures phase diversity. Table~\ref{tab:initial_states} presents these theoretical initial states and the corresponding experimental realizations ($\hat{\rho}_{k}$) obtained via optimized magnetic-field pulse sequences, measured using the QST technique described in Sec.~\ref{sec:QST}. Consistent with the notation of the main text, the columns of the input-state matrix $\mathbb{M}^{(i)}$ are the vectorized density matrices $\sket{\rho_k}$, so that $\mathbb{M}^{(i)}$ serves as the input-state matrix for subsequent evolution.

Each prepared quantum state undergoes unitary and non-unitary evolution under the influence of the target environment and controllable fields. The resulting time-evolved states are reconstructed using QST (see Sec.~\ref{sec:QST}), and their vectorized forms $\sket{\rho'_{k}}$ constitute the output-state matrix $\mathbb{M}^{(o)}$. In the Bloch--Fano representation, the evolution is described by the superoperator $\dhat{P}(t)$:
\begin{equation}
    \mathbb{M}^{(o)} = \dhat{P}(t)\mathbb{M}^{(i)}.
    \label{eq:in_out}
\end{equation}
To address the overcompleteness of the input state set ($N = 15 > d^2 = 9$), we define symmetrized input- and output-state matrices as:
\begin{equation}
    \underline{\mathbb{M}}^{(i,o)} = \mathbb{M}^{(i,o)} \cdot \left( \mathbb{M}^{(i)} \right)^{\mathsf{T}},
\end{equation}
which allows for the reconstruction of the process matrix through linear inversion:
\begin{equation}
    \dhat{P}(t) = \underline{\mathbb{M}}^{(o)} \left( \underline{\mathbb{M}}^{(i)} \right)^{-1}.
\end{equation}

A major limitation of the raw reconstructed process matrix is its susceptibility to state-preparation-and-measurement (SPAM) errors. Because imperfections in state preparation and measurement may partially compensate each other, the reconstruction is subject to a gauge ambiguity that can yield unphysical results. To resolve this, we introduce a zero-time calibration procedure. Under the assumption that the preparation and measurement channels are statistically independent, the reconstructed process matrix at $t=0$ should ideally correspond to the identity superoperator, $\dhat{P}_{\mathrm{ideal}}(0)=\dhat{\mathds{1}}$. Any deviation can be attributed to the bare SPAM contribution, defined by:
\begin{equation}
    \dhat{P}_{\mathrm{SPAM}}=\mathbb{\underline{M}}^{(o)}(0)\left(\mathbb{\underline{M}}^{(i)}\right)^{-1}.
\end{equation}

In practice, direct determination of $\dhat{P}_{\mathrm{SPAM}}$ from a single zero-time process matrix is limited by the experimental inaccessibility of the $t=0$ state and the presence of statistical noise. To mitigate these limitations, we estimate the SPAM contribution through a linear extrapolation procedure. Under the assumption of Markovian dynamics, the process matrix is expressed as:
\begin{equation}
    \dhat{P}_{\mathrm{exp}}(t)=\dhat{P}_{\mathrm{SPAM}}\,\exp\!\left(\dhat{L}\, t\right).
\end{equation}
To estimate the static SPAM contribution, we employ a linearized fitting ansatz in the matrix-logarithm representation:
\begin{equation}
    \log\bigl[\dhat{P}_{\mathrm{exp}}(t)\bigr]\approx \log\bigl(\dhat{P}_{\mathrm{SPAM}}\bigr)+\dhat{L}t,
\end{equation}
which enables the separation of the time-independent offset from the dynamical contribution governed by the Liouvillian. Using a set of process matrices $\{\dhat{P}_{\mathrm{exp}}(t_i)\}$, we determine the optimal estimate of $\dhat{P}_{\mathrm{SPAM}}$ by performing a least-squares fit to this linearized model. Inferring the static SPAM contribution from the time series substantially suppresses the influence of statistical fluctuations compared to a single zero-time measurement.

\begin{figure}[ht]
    \centering
    \includegraphics[width=1\linewidth]{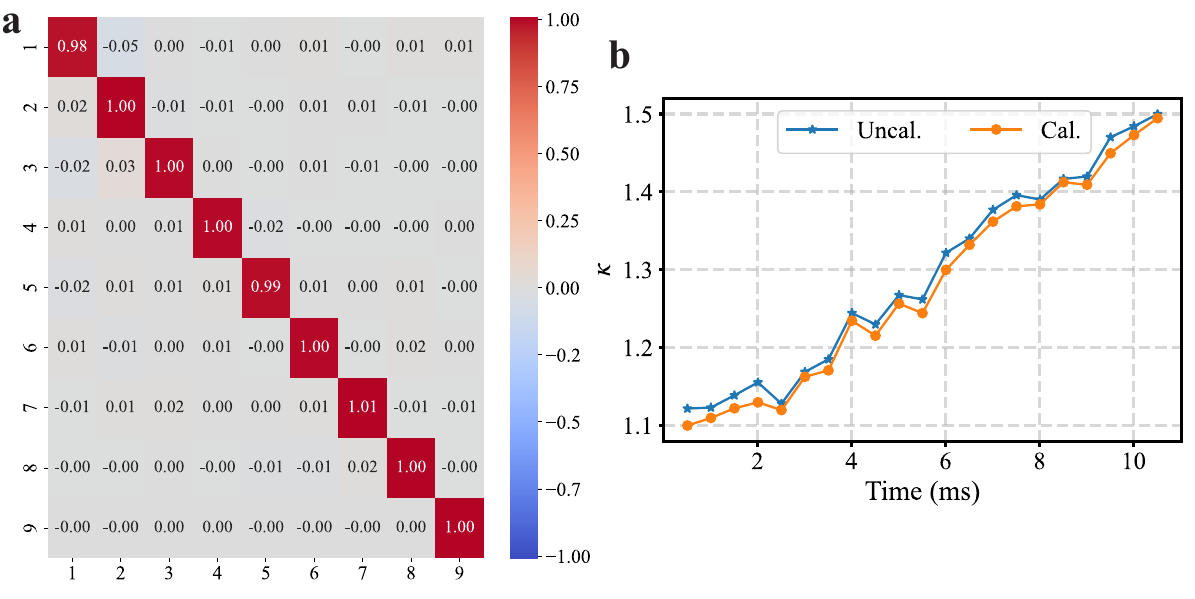}
    \caption{Zero-time calibration of SPAM contributions in the reconstructed process matrices. (a) Estimated calibration matrix $\dhat{P}_{\mathrm{SPAM}}$ obtained from linear extrapolation to zero evolution time. (b) Condition numbers of the time-dependent process matrices before (blue) and after (orange) SPAM calibration. }
    \label{fig:calibration}
\end{figure}

A prerequisite for this procedure is that the matrix logarithm remains uniquely defined. In our experiment, the eigenvalue phases of the reconstructed process matrices are strictly confined to the interval $[-0.35,\,0.35]$ radians, preventing $2\pi$ phase discontinuities and validating the selection of the principal branch. The resulting calibrated matrix $\dhat{P}_{\mathrm{SPAM}}$ is shown in Fig.~\ref{fig:calibration}(a).

Once $\dhat{P}_{\mathrm{SPAM}}$ has been obtained, the measured time-dependent process matrices are corrected according to:
\begin{equation}
    \dhat{P}_{\mathrm{corr}}(t)=\left[\dhat{P}_{\mathrm{SPAM}}\right]^{-1}\cdot \dhat{P}_{\mathrm{exp}}(t),
\end{equation}
isolating the intrinsic system dynamics. The effectiveness of this calibration is quantified in Fig.~\ref{fig:calibration}(b). After calibration, the condition number \cite{Kopciuch2024} is systematically reduced, indicating that the corrected matrices are better conditioned for subsequent inversion and Liouvillian reconstruction.

\begin{figure}[h!]
    \centering
    \includegraphics[width=1\linewidth]{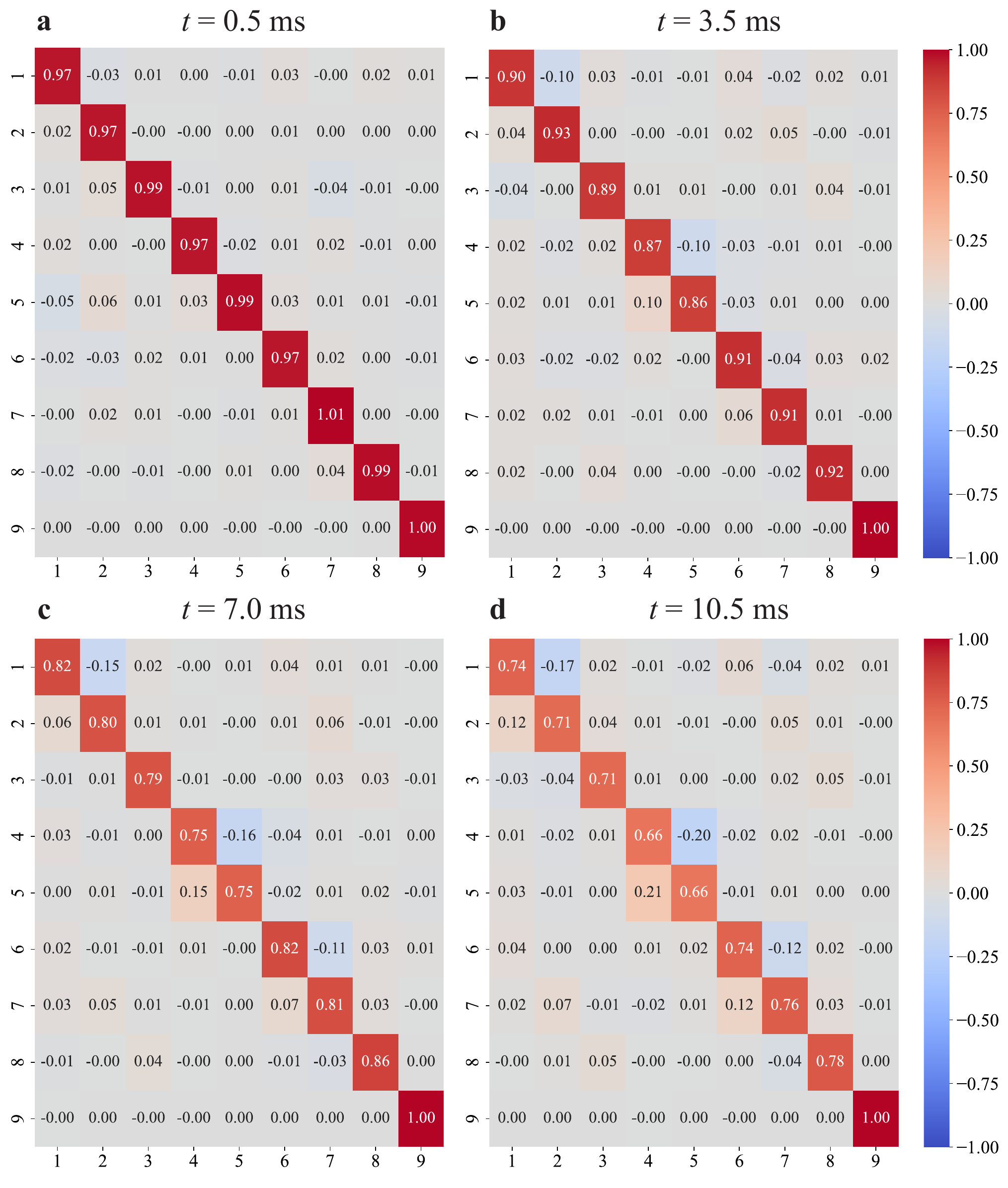}
    \caption{Examples of process matrices $\dhat{P}(t)$ at different evolution times under the action of the effective total relaxation superoperator $\dhat{R}_{T}$.}
    \label{fig:Relaxation Process matrices}
\end{figure}

\begin{figure}[ht]
    \centering
    \includegraphics[width=1\linewidth]{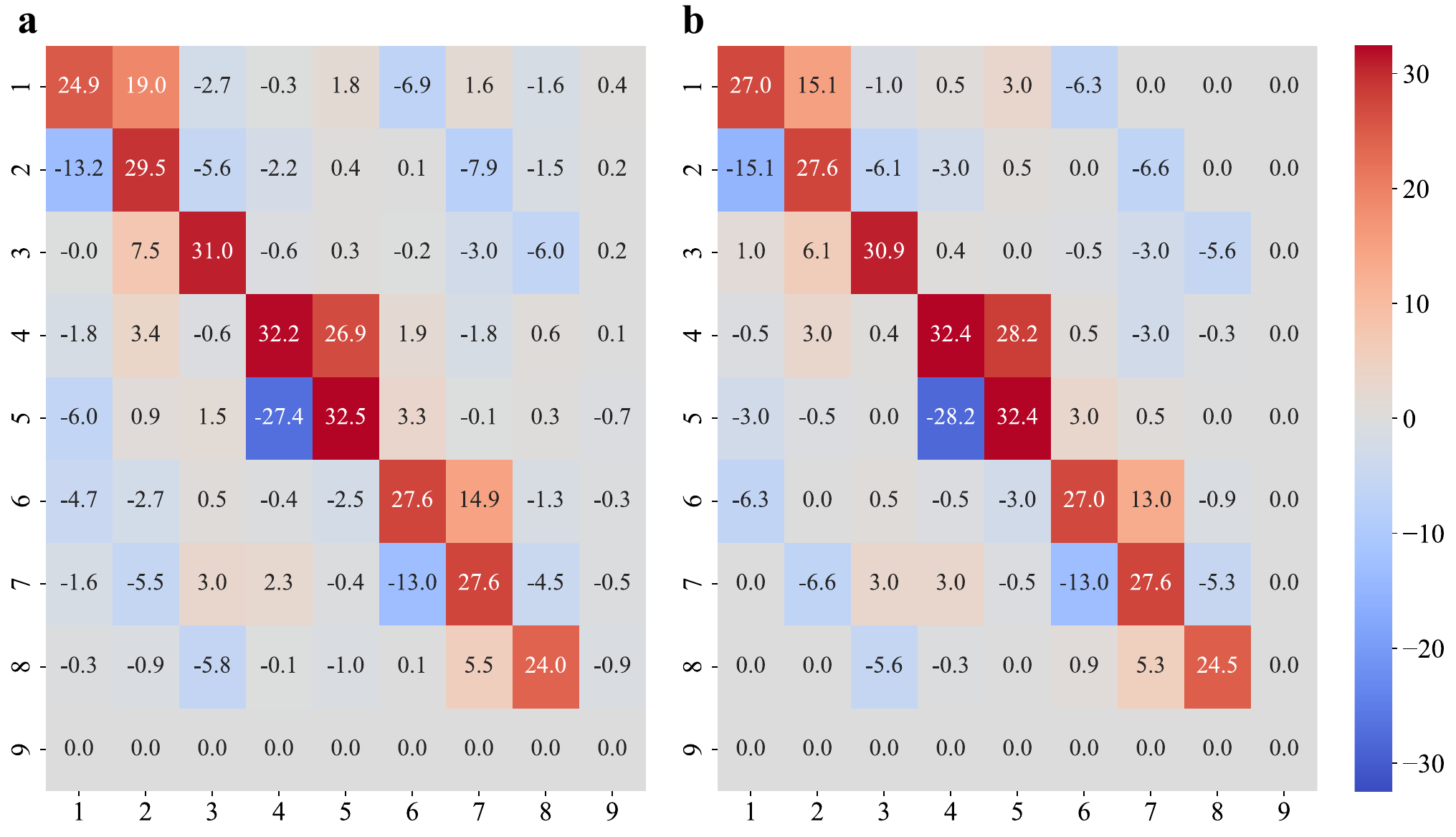}
    \caption{(a) Reconstruction of the effective total relaxation superoperator $\dhat{R}_T$ using the MLE protocol. (b) Estimated relaxation superoperator using a model including isotropic relaxation, residual coherent contribution, and magnetic field gradients.}
    \label{fig:relaxation_estimation}
\end{figure}

Figure~\ref{fig:Relaxation Process matrices} shows examples of raw process matrices in the Bloch--Fano representation, which capture system dynamics driven solely by the effective total relaxation superoperator $\dhat{R}_{T}$. For an ideal physical process, the matrix elements are expected to be 
within the interval $\left[-1, 1\right]$. The calibrated process matrices then enable reconstruction of the dissipator $\dhat{R}_{T}$. After imposing positive 
semi-definiteness on the associated dissipative sector, we obtain the physically admissible dissipator shown in Fig.~\ref{fig:relaxation_estimation}(a). 

To gain further insight into the underlying relaxation mechanisms, we model $\dhat{R}_{T}$ as a sum of three distinct processes: 

\begin{enumerate}
    \item Isotropic relaxation typical for paraffin-coated atomic-vapor cells, represented by $\dhat{R}_{\text{iso}} = \mathds{\hat{1}}_9 - \delta_{99}$, where the last diagonal element remains zero to ensure trace preservation.
    \item Coherent contribution induced by residual magnetic fields together with the AC Stark shift along the $z$-axis arising from leakage of the repump beam, modeled by the Hamiltonian:
    \begin{equation}
        \hat{H}=\sum_{k=x,y,z} (\mathbf{\Omega}_L)_k \hat{F}_k + \Omega_{\mathrm{AC}}\left( 3\hat{F}_z^2 - \hat{\mathbf F}^2 \right).
    \end{equation}    
    \item Dephasing, modeled using three independent Lindblad operators $\hat{L}_k = \gamma_k \hat{F}_k$ along the spatial axes.
\end{enumerate}

From the measured $\dhat{R}_T$ shown in Fig.~\ref{fig:relaxation_estimation}, we estimate the residual Larmor frequencies as $\mathbf{\Omega}_L/2\pi = \left\lbrace -0.6828(85), 0.1142(86), 2.240(34) \right\rbrace\,\si{\hertz}$, the AC Stark-shift frequency as $\mathbf{\Omega}_{\mathrm{AC}}/2\pi = \left\lbrace 0.056(16) \right\rbrace\,\si{\hertz}$, the dephasing rates as $\mathbf{\gamma}_k = \left\lbrace 6.7(12), 6.1(12), 5.60(11) \right\rbrace\,\si{\per\second}$, and the isotropic relaxation rate as $\gamma_i = 14.84(15)\,\si{\per\second}$.

\section{Choi matrix in Bloch--Fano representation \label{sec:Choi}}

The Choi matrix constitutes an isomorphic representation of a quantum channel through the Choi–Jamiolkowski isomorphism, facilitating the verification of complete positivity and the analysis of the properties of reconstructed processes \cite{choi1975completely}. To construct the Choi matrix in the Bloch--Fano representation, we utilize the Choi--Jamiolkowski isomorphism \cite{jamiolkowski1972linear}:
\begin{equation}
    \mathbb{J} = (\mathcal{E} \otimes \mathcal{I}) \dyad{\Omega},
\end{equation}
where $\mathcal{E}$ is the linear map describing the process of interest, $\mathcal{I}$ denotes the identity map on the ancillary subsystem, and $ \ket{\Omega} = 1/\sqrt{d}\sum_{i} \ket{i}\otimes\ket{i} $ is the maximally entangled state. The process $\mathcal{E}$ maps the initial state according to $\hat{\rho}(t)=\mathcal{E}[\hat{\rho}(0)]$.

To derive an explicit relation between the Choi matrix and the process matrix in the
Bloch--Fano basis, we first expand $\dyad{\Omega}$ in the operator basis,$ \dyad{\Omega} = \sum_{\alpha, \beta} E_{\alpha \beta} \hat{\sigma}_{\alpha} \otimes \hat{\sigma}_{\beta}$. The expansion coefficients are obtained by projection:
\begin{equation}
\begin{split}
    E_{\alpha \beta} &= \frac{1}{4}\Tr \Big[ \hat{\sigma}_{\alpha} \otimes \hat{\sigma}_{\beta} \dyad{\Omega} \Big]\\
    &= \frac{1}{4d}\sum_{ij}\Tr \Big[ \hat{\sigma}_{\alpha} \dyad{i}{j}\Big] \Tr \Big[ \hat{\sigma}_{\beta} \dyad{i}{j}\Big]\\
    &= \frac{1}{4d}\sum_{ij} \big(\hat{\sigma}_{\alpha} \big)_{ij} \big(\hat{\sigma}_{\beta} \big)_{ij}\\
    &=  \frac{1}{4d}\sum_{ij} \big(\hat{\sigma}_{\alpha} \big)_{ij} \big(\hat{\sigma}_{\beta}^{\rm T} \big)_{ji} \\
    &= \frac{1}{4d}\Tr \Big[ \hat{\sigma}_{\alpha} \hat{\sigma}_{\beta}^{\rm T} \Big] = \frac{1}{2d}s_{\beta} \delta_{\alpha \beta}.
\end{split}
\end{equation}
Here, $s_{\beta} = \pm 1$ characterizes the parity of the basis operator $\hat{\sigma}_{\beta}^{\rm T}$, namely $\hat{\sigma}_{\beta}^{\rm T}=s_{\beta}\hat{\sigma}_{\beta}$; explicitly, $s_{\beta} = +1$ for symmetric generators and $s_{\beta} = -1$ for antisymmetric generators. Consequently,
\begin{equation}
    \dyad{\Omega} = \frac{1}{2d} \sum_{\alpha,\beta} s_{\beta}\delta_{\alpha\beta}\hat{\sigma}_{\alpha} \otimes \hat{\sigma}_{\beta}=\frac{1}{2d} \sum_{\alpha} \hat{\sigma}_{\alpha} \otimes \hat{\sigma}_{\alpha}^{\rm T}.
    \label{eq: MES}
\end{equation}
Having established the expansion of  $\dyad{\Omega}$ in the Bloch--Fano operator basis, we now represent the input and output density operators in the same basis:
\begin{equation}
    \hat{\rho}(0) = \sum_{\alpha}\rho_{\alpha}\hat{\sigma}_{\alpha},\quad\hat{\rho}(t)=\mathcal{E}[\hat{\rho}(0)]=\sum_{\alpha}\rho_{\alpha}\mathcal{E}(\hat{\sigma}_{\alpha})=\sum_{\beta}\rho_{\beta}(t)\hat{\sigma}_{\beta}.
    \label{eq: states}
\end{equation}
In the vectorized representation, the process matrix $\dhat{P}(t)$ is defined as as the linear operator satisfying:
\begin{equation}
    \sket{\rho(t)} = \dhat{P}(t)\sket{\rho(0)},
\end{equation}
which is equivalent to 
\begin{equation}
    \rho_{\beta}(t) = \sum_{\alpha}P_{\beta\alpha}(t)\rho_{\alpha}.
    \label{eq: linear map}
\end{equation}
Comparison of Eq. \eqref{eq: linear map} and Eq. \eqref{eq: states} shows that the process acts on the operator basis as
\begin{equation}
    \label{eq: linear map}
    \mathcal{E}(\hat{\sigma}_\alpha)
    =
    \sum_\beta P_{\beta\alpha}(t)\hat{\sigma}_\beta,
\end{equation}
Substituting Eqs. \eqref{eq: MES} and \eqref{eq: linear map} into the definition of $\mathbb{J}$ yields
\begin{equation}
\begin{split}
    \mathbb{J} &= \big(  \mathcal{E} \otimes \mathcal{I} \big)  \dyad{\Omega} \\
    &= \frac{1}{2d}\sum_{\alpha} \mathcal{E}(\hat{\sigma}_{\alpha} \big) \otimes \mathcal{I}(\hat{\sigma}_{\alpha}^{\rm T}) \\
    &= \frac{1}{2d}\sum_{\alpha \beta} \big( P_{\beta \alpha }(t) \hat{\sigma}_{\beta} \big) \otimes \hat{\sigma}_{\alpha}^{\rm T} \\
    &= \frac{1}{2d}\sum_{\alpha \beta} P_{\beta\alpha}(t) \big( \hat{\sigma}_{\beta} \otimes \hat{\sigma}_{\alpha}^{\rm T} \big).
\end{split}
\end{equation}
This gives the explicit relation between the Choi matrix and the process matrix in the Bloch--Fano basis. The inverse relation follows by projection:
\begin{equation}
\begin{split}
    \Tr \Big[ \mathbb{J} \big( \hat{\sigma}_{\beta} \otimes \hat{\sigma}_{\alpha}^{\rm T} \big) \Big] &= \frac{1}{2d}\sum_{\mu \nu} P_{\mu \nu}(t)
       \Tr\big[ \big(\hat{\sigma}_{\nu} \otimes \hat{\sigma}_{\mu}^{\rm T}\big)
                \big( \hat{\sigma}_{\beta} \otimes \hat{\sigma}_{\alpha}^{\rm T} \big) \big] \\
    &= \frac{2}{d} P_{\beta \alpha}(t).
\end{split}
\end{equation}
Thus, the Choi matrix and the process matrix are related by a linear, invertible transformation and contain the same dynamical information.

\section{Spectral Analysis of the Choi Matrix in Dissipation}

\begin{figure}[ht]
    \centering
    \includegraphics[width=0.75\linewidth]{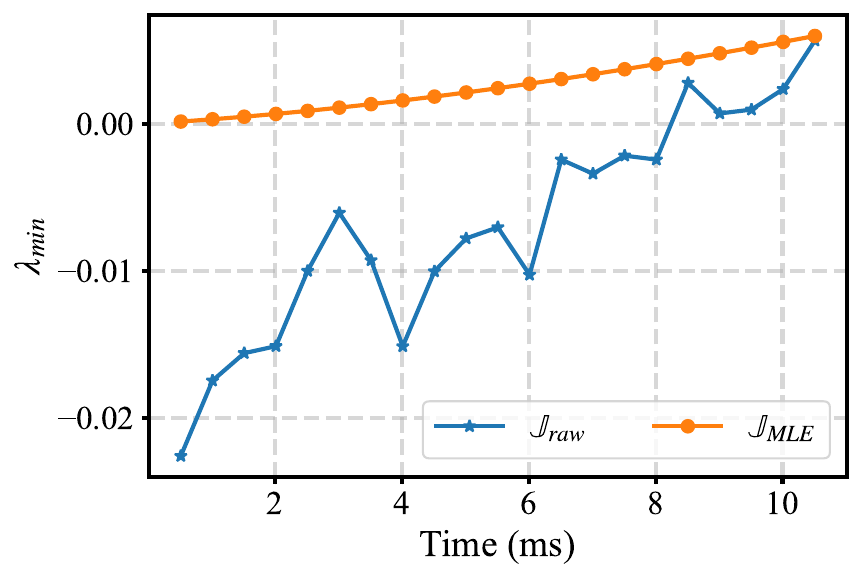}
    \caption{Evolution of the most negative eigenvalue ($\lambda_{\min}$) of the Choi matrix under the influence of dissipation. The blue and orange curves are obtained from the raw experimental process matrix and the process matrix reconstructed via MLE, respectively.}
    \label{fig:CPTP}
\end{figure}

As discussed in the main text, direct linear algebraic inversion of the state matrices does not in general guarantee a completely positive map in the presence of finite sampling noise and residual SPAM errors. As a result, the raw algebraic estimator may lie outside the physical completely positive and trace-preserving (CPTP) set before the enforcement of physical constraints, and the subsequent projection onto the positive semi-definite cone introduces a corresponding systematic correction.

To rigorously quantify the extent of this deviation, we analyze the eigenspectrum of the unconstrained Choi matrix $\mathbb{J}(t)$. As shown in Fig.~\ref{fig:CPTP}, we monitor its smallest eigenvalue, $\lambda_{\min}$, which provides a direct indicator of deviations from complete positivity. For the Choi matrix derived from the raw process matrix, $\lambda_{\min}$ exhibits a small negative value at short evolution times, but increases towards non-negative values at later times. Given its small magnitude, this behavior is most naturally attributed to short-time reconstruction uncertainty rather than to a genuine physical feature of the dissipative dynamics. By contrast, for the regularized Maximum Likelihood Estimation (MLE) reconstruction, the smallest eigenvalue remains non-negative over the full time range and increases monotonically with evolution time. This demonstrates that the physically constrained reconstruction suppresses the spurious short-time violation of complete positivity present in the raw inversion and yields a process matrix consistent with a physically admissible GKLS description.

\section{Uncertainty characterization of QPT}

\begin{figure}[htbp]
    \centering
    \includegraphics[width=1\linewidth]{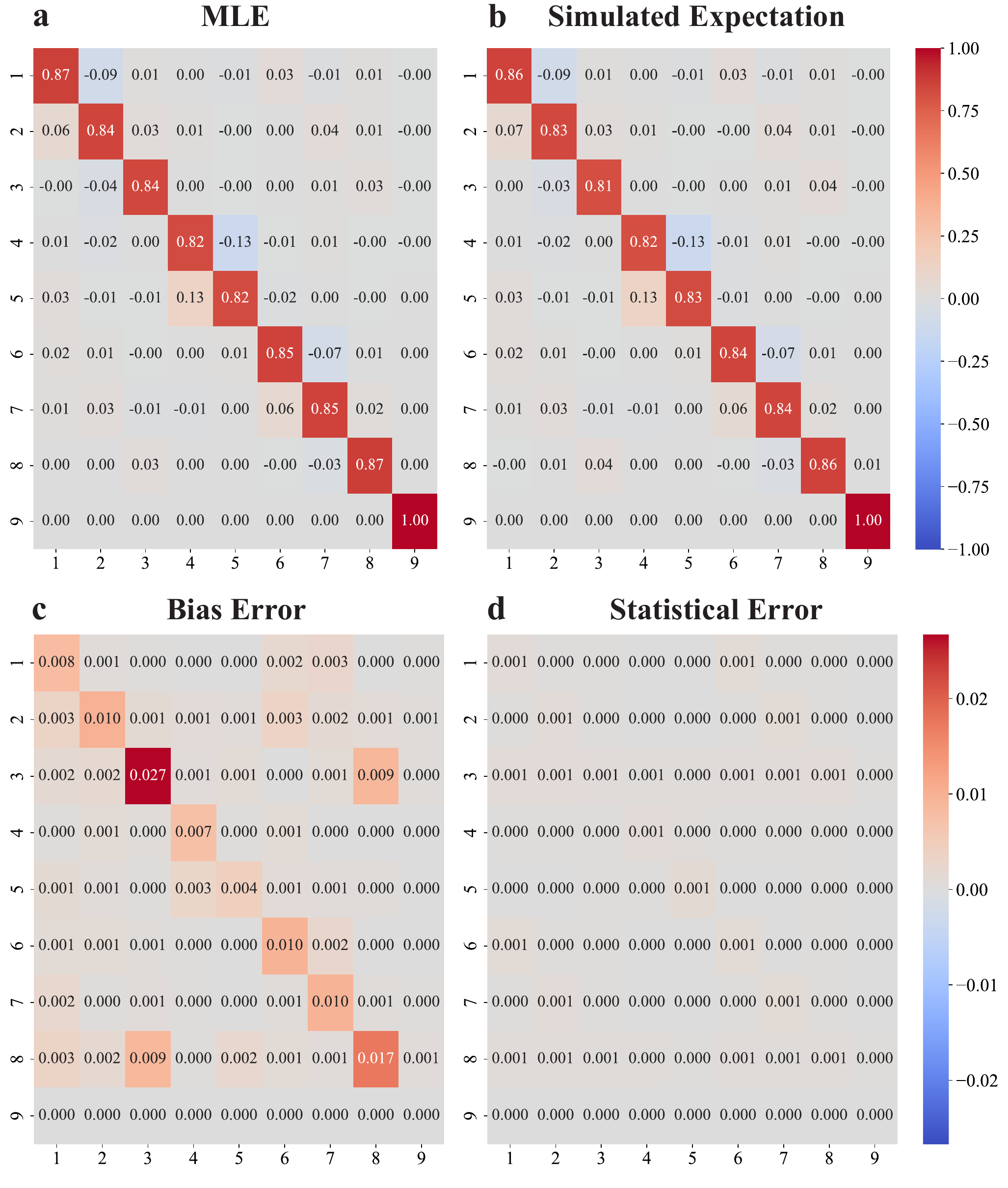}
    \caption{(a) Process matrix $\dhat{P}(t)$ reconstructed via MLE at $t=\SI{5.5}{\milli\second}$ for the effective total relaxation superoperator $\dhat{R}_{T}$. (b) Ensemble-averaged process matrix derived from Monte Carlo resampling. (c) Bias error matrix quantifying the deviation between the MLE reconstruction and the simulated expectation. (d) Statistical uncertainty matrix representing the standard deviation across the simulated process matrices.}
    \label{fig:uncertainty}
\end{figure}

Two complementary approaches exist to estimate experimental uncertainty. The first follows the analytical law of uncertainty propagation, where the uncertainty of the output is obtained by weighting the uncertainties of the input variables with first-order partial derivatives. Although computationally efficient, this method assumes local linearity. Since data processing involves highly nonlinear transformations, such as matrix inversion and matrix logarithm, a first-order approximation is insufficient to capture the full uncertainty budget.

To circumvent these limitations, we implemented the second approach, i.e., Monte Carlo resampling method (parametric bootstrapping) \cite{schmied2016quantum}. Instead of relying on local derivatives, this method propagates uncertainties by repeatedly resampling the input and output state vectors from their experimental probability distributions to generate a statistical ensemble of process matrices. 

The QST procedure provides an estimate of the expected state vector $\sket{\bar{\rho}}$ and its uncertainties $u(\sket{\rho}_i)$ for all quantum states. Due to the trace-preservation constraint, the ninth component is fixed to $\sket{\bar{\rho}}_9 = \sqrt{1/6}$. Consequently, its statistical uncertainty vanishes ($u(\sket{\rho}_9) = 0$). We generate 1000 synthetic data sets by resampling each component of $\sket{\rho}$ from a normal distribution centered on its expectation value. For each dataset, a process matrix is reconstructed, and the associated Liouvillian superoperator is extracted. The uncertainties are then determined by statistical analysis of the resulting ensemble.

Figure~\ref{fig:uncertainty} shows the reconstructed process matrix at $t=\SI{5.5}{\milli\second}$, the ensemble-averaged process matrix generated from Monte Carlo simulations, the resulting bias error matrix, and the statistical uncertainty matrix. In principle, the MLE-based reconstructed process matrix should fall within the confidence intervals defined by the simulated expectation values. However, Figs.~\ref{fig:uncertainty}(c) and (d) show that the magnitudes of the bias error matrix generally exceed those of the statistical error matrix. This discrepancy is a feature of constrained MLE in quantum tomography \cite{schwemmer2015systematic}. Because experimental noise frequently scatters unconstrained estimates outside the physically allowed parameter space, the MLE algorithm projects these results back onto the boundary of the CPTP cone. This asymmetric truncation inherently skews the distribution of the reconstructed parameters, introducing a systematic algorithmic bias. We treat the observed bias between the primary MLE reconstruction and the Monte Carlo expectation as a systematic uncertainty component, $\Delta_{\mathrm{bias}}$, combining it with the statistical variance in quadrature:
\begin{equation}
    u_{\mathrm{total}}=\sqrt{\Delta^{2}_{\mathrm{bias}}+u^2_{\mathrm{stat}}}~,
\end{equation}
where $u_{\mathrm{stat}}$ denotes the Monte Carlo ensemble standard deviation. The treatment of the algorithmic bias as a systematic uncertainty is applied to the evaluation of the process matrix, the Liouvillian superoperators, and all derived experimental parameters.

\section{Hamiltonian reconstruction\label{sec:estimators}}

The fundamental approach for estimating the underlying generator of a quantum process is through direct algebraic inversion. Provided the Liouvillian $\dhat{L}$ is time-independent over the interrogation time $t$, the measured process matrix is reconstructed using the matrix logarithm:
\begin{equation}
    \dhat{L} = \dfrac{1}{t} \log(\dhat{P}(t)) = \dfrac{1}{t} \log\left[\underline{\mathbb{M}}^{(o)}\left(\underline{\mathbb{M}}^{(i)}\right)^{-1}\right],
\end{equation}
where $\underline{\mathbb{M}}^{(i)}$ and $\underline{\mathbb{M}}^{(o)}$ are the symmetrized matrices containing the input and the output states of the tomography protocol. 

Before extracting the control Hamiltonian, we subtract the effective relaxation superoperator, obtained as described previously:  
\begin{equation}
   \dhat{H}_{C}= i\left(\dhat{L}+\dhat{R}_{T}\right).
\end{equation}

Due to finite sampling and experimental noise, the directly reconstructed superoperator $\dhat{H}_{C}$ is an a priori unconstrained real matrix, which may not correspond to a strictly Hermitian Hamiltonian operator. To extract the closest physical Hamiltonian, we must enforce Hermiticity by projecting this result onto the space of valid Hamiltonian superoperators. 

To construct this mapping, we recall that any Hermitian operator can be expanded in the generalized Pauli basis as $\hat{H} = \sum_k h_k \hat{\sigma}_k$. The algebra of this basis is characterized by the totally antisymmetric structure constants $f_{ijk}$ of $\mathrm{SU}(d)$, defined through the commutation
relation $[\hat{\sigma}_k, \hat{\sigma}_j] = 2i\sum_l f_{kjl}\hat{\sigma}_l$ are given in Eq. \eqref{eq:Hamiltonian_as_dissipator}. This relation directly yields the analytical mapping between the Hamiltonian operator $\hat{H}$ and its superoperator representation $\dhat{H}$:
\begin{equation}
    \begin{aligned}
    \label{eq: Hamiltonian Conversion}
        \dhat{H}_{ij} &= \dfrac{1}{2} \Tr(\comm{\hat{H}}{\hat{\sigma}_j}\hat{\sigma}_i) = \frac{1}{2}\Tr \left( \sum_{k}h_{k}\comm{\hat{\sigma}_{k}}{\hat{\sigma}_{j}}\hat{\sigma}_{i} \right)\\
        &= \frac{1}{2} \sum_{k}h_{k}\Tr \left( 2i\sum_{l}f_{kjl}\hat{\sigma}_{l}\hat{\sigma}_{i} \right)\\
        &= 2i\sum_{k}h_{k}f_{kji},
    \end{aligned}
\end{equation}
where we have used the orthogonality relation $\Tr(\hat{\sigma}_l\hat{\sigma}_i) = 2\delta_{li}$.

For the specific case of a qutrit, we can explicitly parametrize a general $3\times3$ Hermitian operator $\hat{H}$ using nine real parameters $\left\{H_{k}\right\}_{k=1}^{9}$:
\begin{equation}
\hat{H}=
\begin{pmatrix}
 H_{1}            & H_{2}-iH_{3} & H_{4}-iH_{5} \\[2pt]
 H_{2}+iH_{3}     & H_{6}        & H_{7}-iH_{8} \\[2pt]
 H_{4}+iH_{5}     & H_{7}+iH_{8} & H_{9}
\end{pmatrix}.
\label{eq:Hamiltonian_param}
\end{equation}

By evaluating the commutation relations exhaustively over the Gell-Mann indices and mapping the coefficients back to the standard $3\times3$ Hermitian matrix entries $\{H_k\}$, we systematically derive the explicit theoretical superoperator structure:
\begin{widetext}
\begin{equation}
    \dhat{H}_{\mathrm{th}} = -i\left(
\begin{array}{ccccccccc}
 0 & H_1-H_6 & -2 H_3 & H_8 & -H_7 & H_5 & -H_4 & 0 & 0 \\
 H_6-H_1 & 0 & 2 H_2 & H_7 & H_8 & -H_4 & -H_5 & 0 & 0 \\
 2 H_3 & -2 H_2 & 0 & H_5 & -H_4 & -H_8 & H_7 & 0 & 0 \\
 -H_8 & -H_7 & -H_5 & 0 & H_1-H_9 & H_3 & H_2 & -\sqrt{3} H_5 & 0 \\
 H_7 & -H_8 & H_4 & H_9-H_1 & 0 & -H_2 & H_3 & \sqrt{3} H_4 & 0 \\
 -H_5 & H_4 & H_8 & -H_3 & H_2 & 0 & H_6-H_9 & -\sqrt{3} H_8 & 0 \\
 H_4 & H_5 & -H_7 & -H_2 & -H_3 & H_9-H_6 & 0 & \sqrt{3} H_7 & 0 \\
 0 & 0 & 0 & \sqrt{3} H_5 & -\sqrt{3} H_4 & \sqrt{3} H_8 & -\sqrt{3} H_7 & 0 & 0 \\
 0 & 0 & 0 & 0 & 0 & 0 & 0 & 0 & 0 \\
\end{array}
\right).
\label{eq:H_superoperator}
\end{equation}
\end{widetext}

Finally, we equate the matrix elements of this theoretical form with those of the directly reconstructed experimental superoperator, $\left(\dhat{H}_{\mathrm{th}}\right)_{ij} = \left( \dhat{H}_C(t) \right)_{ij}$, obtaining an overdetermined linear system for the parameters $H_{k}$. Solving this system in the least-squares sense yields the physical Hermitian Hamiltonian that is most consistent with the measured process.

For time-dependent reconstructions described in the main text, this algebraic inversion and subsequent least-squares mapping onto the physical parameters $\{H_k(t_n)\}$ is performed independently for each coarse-grained time step $\Delta t$, which avoids cumulative numerical integration errors. For multiparameter estimation, these extracted discrete parameters are subsequently projected onto specific continuous analytical waveforms $\Omega_{k}(t)$.

\section{Reconstruction of static magnetic fields}
\begin{figure}[htbp]
    \centering
    \includegraphics[width=1\linewidth]{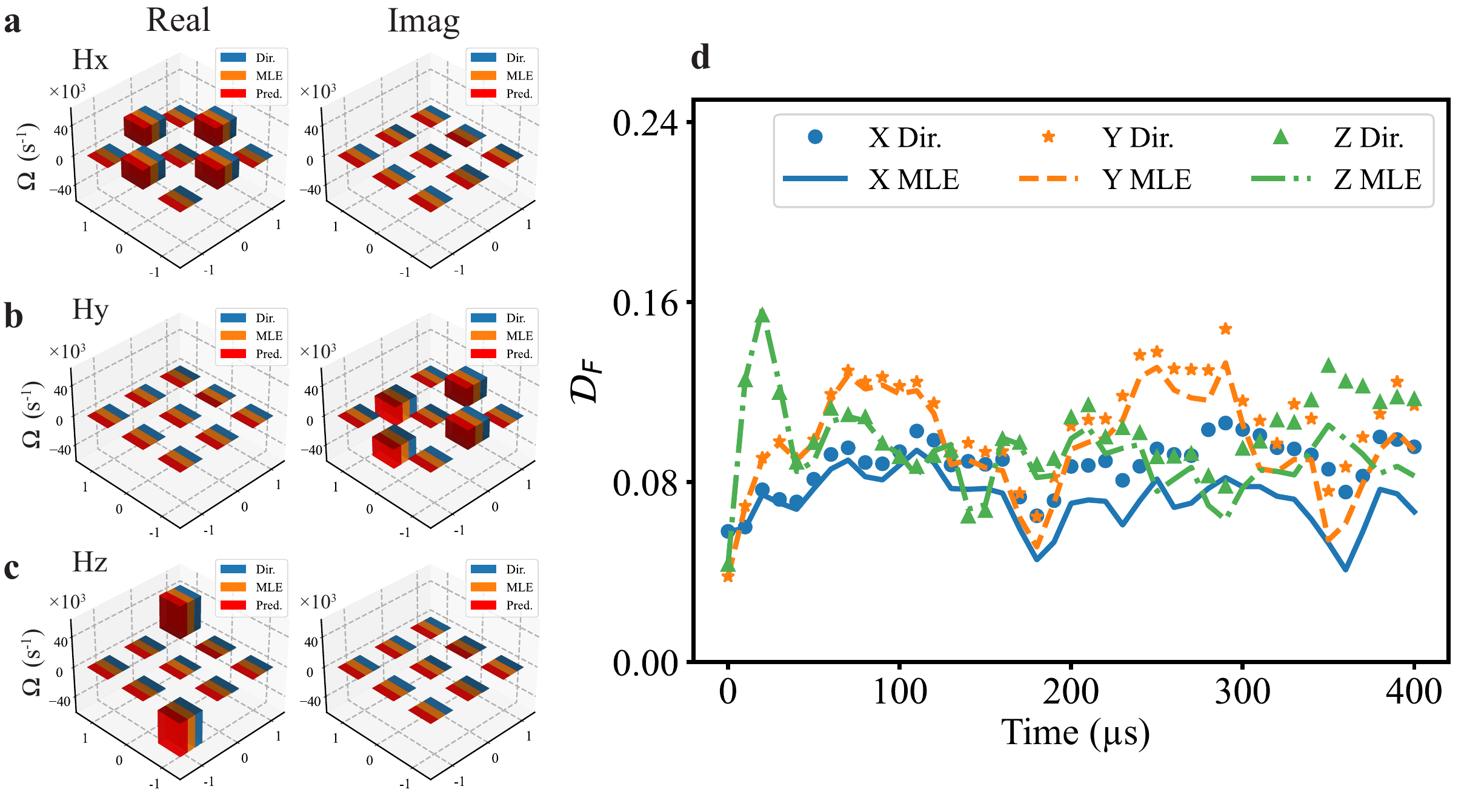}
    \caption{(a)-(c) Comparison of the Hamiltonian operators. When the magnetic field is aligned along the $x$-, $y$-, and $z$-axes, the relative errors via MLE are $\mathcal{D}_{F}=0.01835(37)$, $0.0219(11)$, and $0.02853(15)$, respectively, while for direct reconstruction they are $\mathcal{D}_{F}= 0.0186(23)$, $0.0212(19)$, and $0.0439(18)$, respectively. (d) Normalized Frobenius distance between the experimentally measured process superoperators and those generated from the estimated $\dhat{H}_{C}$ using direct reconstruction (points) and MLE (lines) when the static magnetic field is applied along the $x$-, $y$-, and $z$-axes.}
    \label{fig:linear_magnetic_field}
\end{figure}
As a metrological validation of our time-independent Hamiltonian reconstruction, we evaluate the reconstruction of the Hamiltonian associated with the linear Zeeman effect. Figures~\ref{fig:linear_magnetic_field}(a)--(c) compare the theoretical predictions, direct reconstruction, and MLE for magnetic fields aligned with the $x$-, $y$-, and $z$-axes, respectively. Quantitative analysis shows that the relative errors between the theoretical Hamiltonian superoperator and the reconstruction via MLE are $0.01835(37)$, $0.0219(11)$, and $0.02853(15)$, while those for direct reconstruction are $0.0186(23)$, $0.0212(19)$, and $0.0439(18)$, respectively. 

Figure~\ref{fig:linear_magnetic_field}(d) presents a quantitative relative error analysis comparing the experimentally measured process matrices with the process matrices $\dhat{P}$ generated from the estimated Hamiltonians. All relative errors are bounded by $0.158(65)$. The close agreement between reconstructions via both methods and predictions under static magnetic field conditions confirms the reliability of the experimental system and the Hamiltonian reconstruction method.

\begin{longtable}[ht]{cccc}
  \caption{Initial states: Theory vs. Experiment}
  \label{tab:initial_states} \\
  \toprule
  \textbf{States} & \textbf{Theory} & \textbf{Experiment}&\textbf{Fidelity} \\
  \midrule
  \endfirsthead 
  \multicolumn{4}{c}{Table \ref{tab:initial_states} continued} \\
  \toprule
  \textbf{States} & \textbf{Theory} & \textbf{Experiment}&\textbf{Fidelity} \\
  \midrule
  \endhead %
  \bottomrule
  \endfoot %
  \bottomrule
  \endlastfoot %

  $\hat{\rho}_1$ & $\begin{pmatrix} 1 & 0 & 0 \\ 0 & 0 & 0 \\ 0 & 0 & 0 \end{pmatrix}$ & $\begin{pmatrix} 0.919& 0.001-0.080i&0.011+0.011i \\ 0.001+0.080i & 0.045 & -0.012-0.017i \\ 0.011-0.011i & -0.012+0.017i & 0.036 \end{pmatrix}$ & 0.919 \\
    \\
    $\hat{\rho}_2$ & $\begin{pmatrix} 0 & 0 & 0 \\ 0 & 1 & 0 \\ 0 & 0 & 0 \end{pmatrix}$ & $\begin{pmatrix} 0.069& -0.017+0.039i& -0.018-0.001i \\ -0.017-0.039i& 0.878& 0.004+0.007i \\ -0.018+0.001i& 0.004-0.007i& 0.053 \end{pmatrix}$ & 0.878\\
    \\
    
    $\hat{\rho}_3$ & $\begin{pmatrix} 0 & 0 & 0 \\ 0 & 0 & 0 \\ 0 & 0 & 1 \end{pmatrix}$ & $\begin{pmatrix} 0.036& 0.011+0.015i& 0.007-0.027i \\ 0.011-0.015i& 0.072& -0.038i \\ 0.007+0.027i& 0.038i& 0.892 \end{pmatrix}$ & 0.892 \\
    \\
    $\hat{\rho}_4$ & $\begin{pmatrix} 0.5 & 0.5 & 0 \\ 0.5 & 0.5 & 0 \\ 0 & 0 & 0 \end{pmatrix}$ & $\begin{pmatrix} 0.460& 0.423-0.018i& 0.020-0.030i \\ 0.423+0.018i& 0.478& 0.044-0.077i \\ 0.020+0.030i& 0.044+0.077i& 0.062 \end{pmatrix}$ & 0.892 \\
    \\
    $\hat{\rho}_5$ & $\begin{pmatrix} 0.5 & -0.5i & 0 \\ 0.5i & 0.5 & 0 \\ 0 & 0 & 0 \end{pmatrix}$ & $\begin{pmatrix} 0.467& 0.041-0.436i& -0.010+0.021i \\ 0.041+0.436i& 0.480& -0.035-0.033i \\ -0.010-0.021i& -0.035+0.033i& 0.053 \end{pmatrix}$ & 0.909 \\
    \\
    $\hat{\rho}_6$ & $\begin{pmatrix} 0.5 & -0.5 & 0 \\ -0.5 & 0.5 & 0 \\ 0 & 0 & 0 \end{pmatrix}$ & $\begin{pmatrix} 0.473& -0.437-0.092i& 0.047-0.026i \\ -0.437+0.092i& 0.514& -0.043+0.045i \\ 0.047+0.026i&-0.043-0.045i& 0.013 \end{pmatrix}$ & 0.931 \\
    \\
    $\hat{\rho}_7$ & $\begin{pmatrix} 0.5 & 0.5i & 0 \\ -0.5i & 0.5 & 0 \\ 0 & 0 & 0 \end{pmatrix}$ & $\begin{pmatrix} 0.472& -0.144+0.402i& -0.053+0.036i \\ -0.144-0.402i& 0.487& 0.023+0.021i \\ -0.053-0.036i& 0.023-0.021i& 0.041 \end{pmatrix}$ & 0.882 \\
    \\
    $\hat{\rho}_8$ & $\begin{pmatrix} 0 & 0 & 0 \\ 0 & 0.5 & 0.5 \\ 0 & 0.5 & 0.5 \end{pmatrix}$ & $\begin{pmatrix} 0.034& 0.066+0.011i& 0.042-0.031i \\ 0.066-0.011i& 0.532& 0.410-0.075i \\ 0.042+0.031i& 0.410+0.075i& 0.434 \end{pmatrix}$ & 0.893 \\
    \\
    $\hat{\rho}_9$ & $\begin{pmatrix} 0 & 0 & 0 \\ 0 & 0.5 & -0.5i \\ 0 & 0.5i & 0.5 \end{pmatrix}$ & $\begin{pmatrix} 0.041& 0.025-0.052i& -0.047+0.008i \\ 0.025+0.052i& 0.544& -0.033-0.421i \\ -0.047-0.008i& -0.033+0.421i& 0.415 \end{pmatrix}$ & 0.901 \\
    \\
    $\hat{\rho}_{10}$ & $\begin{pmatrix} 0 & 0 & 0 \\ 0 & 0.5 & -0.5 \\ 0 & -0.5 & 0.5 \end{pmatrix}$ & $\begin{pmatrix} 0.054& -0.066-0.019i& 0.027+0.005i \\ -0.066+0.019i& 0.513& -0.408+0.006i \\ 0.027-0.005i& -0.408-0.006i& 0.433 \end{pmatrix}$ & 0.881 \\    
    \\
    $\hat{\rho}_{11}$ & $\begin{pmatrix} 0 & 0 & 0 \\ 0 & 0.5 & 0.5i \\ 0 & -0.5i & 0.5 \end{pmatrix}$ & $\begin{pmatrix} 0.057& -0.049+0.068i& -0.022+0.005i \\ -0.049-0.068i& 0.527& -0.038+0.419i \\ -0.022-0.005i& -0.038-0.419i& 0.416 \end{pmatrix}$ & 0.890 \\
    \\
    $\hat{\rho}_{12}$ & $\begin{pmatrix} 0.5 & 0 & 0.5 \\ 0 & 0 & 0 \\ 0.5 & 0 & 0.5 \end{pmatrix}$ & $\begin{pmatrix} 0.523& -0.016-0.010i& 0.415-0.002i \\ -0.016+0.010i& 0.037& -0.007-0.002i \\ 0.415+0.002i& -0.007+0.002i& 0.440 \end{pmatrix}$ & 0.897 \\
    \\
    $\hat{\rho}_{13}$ & $\begin{pmatrix} 0.5 & 0 & -0.5i \\ 0 & 0 & 0 \\ 0.5i & 0 & 0.5 \end{pmatrix}$ & $\begin{pmatrix} 0.540& -0.041-0.010i& 0.063-0.435i \\ -0.041+0.010i& 0.036& 0.021+0.025i  \\ 0.063+0.435i& 0.021-0.025i& 0.424 \end{pmatrix}$ & 0.917 \\
    \\
    $\hat{\rho}_{14}$ & $\begin{pmatrix} 0.5 & 0 & -0.5 \\ 0 & 0 & 0 \\ -0.5 & 0 & 0.5 \end{pmatrix}$ & $\begin{pmatrix} 0.540& -0.005+0.013i& -0.397-0.070i \\ -0.005-0.013i& 0.025& 0.012+0.007i  \\ -0.397+0.070i& 0.012-0.007i& 0.435 \end{pmatrix}$ & 0.885 \\
    \\
    $\hat{\rho}_{15}$ & $\begin{pmatrix} 0.5 & 0 & 0.5i \\ 0 & 0 & 0 \\ -0.5i & 0 & 0.5 \end{pmatrix}$ & $\begin{pmatrix} 0.525& 0.019+0.022i& -0.134+0.404i \\ 0.019-0.022i& 0.044& 0.004-0.006i  \\ -0.134-0.404i& 0.004+0.006i& 0.431 \end{pmatrix}$ & 0.882 
\end{longtable}

\bibliography{bibliography.bib}